\documentclass[12pt]{article}
\usepackage[english]{babel}
\usepackage{amsmath}
\usepackage{amssymb}
\usepackage{amsfonts}
\usepackage{amsthm}
\usepackage{graphics,graphicx}
\usepackage{color}
\usepackage{epstopdf}

\theoremstyle{plain}

\newtheorem{theorem}{Theorem}
\newtheorem*{proposition*}{Proposition}

\newtheorem*{corollary*}{Corollary}
\newtheorem*{Corollary*}{Important corollary}

\theoremstyle{definition}

\theoremstyle{remark}

\newtheorem*{remark*}{Remark}

\newtheorem*{example*}{Example}

\newtheorem*{examples*}{Examples}

\numberwithin{equation}{section}
\numberwithin{lemma}{section}
\numberwithin{theorem}{section}
\numberwithin{hypothesis}{section}
\numberwithin{definition}{section}
\numberwithin{example}{section}
\numberwithin{corollary}{section}
\numberwithin{remark}{section}
\begin{document}
\begin{center}
{\LARGE Moving Stone in the Hele-Shaw Flow}

\vspace{3mm}
{\bf Gennady Mishuris$^1$, Sergei Rogosin$^{1, 2, *}$, Michal Wrobel$^1$}
\end{center}

\vspace{3mm}
{\footnotesize\it $^1$Aberystwyth University, Penglais, SY23 3BZ Aberystwyth, UK;

\vspace{1mm}
e-mails: ggm@aber.ac.uk; ser14@aber.ac.uk; miw15@aber.ac.uk}

\vspace{3mm}
{\footnotesize\it $^2$Belarusian State University, Nezavisimosti Ave., 4, 220030 Minsk, Belarus;

\vspace{1mm}
e-mail: rogosin@bsu.by}


%\maketitle

\begin{abstract}
\noindent Asymptotic analysis of the Hele-Shaw flow with a small
moving obstacle is performed. The method of solution utilises the
uniform asymptotic formulas for Green's and Neumann functions recently
obtained by V.~Maz'ya and A.~Movchan.
Theoretical results of the paper are
illustrated by the numerical simulations.
% New features of the
%flows in the Hele-Shaw cell are discovered.

\vspace{2mm}

\vspace{2mm} \noindent{\it AMS 2010 Mathematics Subject
Classification}: 76D27, 35J25, 35R35

\end{abstract}

\noindent $^{*}${\footnotesize{Corresponding author.}}

\date{\today}

%
%
%
%\maketitle
%
%
%\dedication

%

\section{Introduction}

{The paper is devoted to the asymptotic study of the flow in the Hele-Shaw cell with presence 
of moving obstacle in the flow.}

The Hele-Shaw problem (\cite{HeleS}) deals with the
description of the free boundary encircling the domain occupied by
incompressible fluid in the so called Hele-Shaw cell  (see, e.g.
\cite{GustVas}, \cite{Vasil}), i.e. in a narrow space between two parallel plates. Different driving mechanisms can be considered for the
fluid flow, e.g. presence of a source/sink in the fluid domain.

Various physical assumptions lead to different formulations of the respective boundary value
problems. A comprehensive discussion on this topic can be found in the recent book by Gustafsson and Vasil'ev
\cite{GustVas}.

There exist two basic mathematical models for the flow in
the Hele-Shaw cell. The {\it complex-analytic model} is formulated
as a nonlinear mixed boundary value problem with respect to a
family of conformal mappings of the canonical domain onto the domain
occupied by the fluid. This approach goes back to the work by
Po\-lu\-ba\-ri\-no\-va-Kochina \cite{P-K} and Galin \cite{Galin}. The proof
of the existence (locally in time) and uniqueness of analytic
solutions to this model was done by Kufarev $\&$ Vinogradov
\cite{VinKu} (rediscovered later by Richardson \cite{Rich72}) on
the  basis of the method of successive approximations. Simplified
proof of existence and uniqueness of an analytic solution was
given by Reissig $\&$ von Wolfersdorf \cite{ReivW93}
(see also \cite{ReiMN93}, \cite{ReiRog}). In this work
the model was interpreted as a special case of an abstract
Cauchy-Kovalevsky problem, which was solved by a variant of
the Cauchy-Kovalevsky theorem (\cite{Nir},
\cite{Nish}, \cite{Ovsja}). See also \cite{DalMcC} and references
therein for the survey of recent results on the complex-analytic Hele-Shaw
problem in doubly connected domains. Let us note that there are some similarities
between movement of the rigid body and movement of bubbles
in the flow (for the discussion of the latter process
we refer, e.g., to the article \cite{EntEti11} and references therein).

{In our study we use} the {\it real-variable model} proposed by Gustafsson
\cite{Gust84} the flow is described by a family of
parametrizations of the boundary of fluid domain (see also \cite{HohRei}).
This model was
generalized to multi-dimensional case by Begehr $\&$ Gilbert
\cite{BegGilb}.
 Among variants of the proof of existence and
uniqueness for this model we have to point out the papers by
Reissig \cite{ReiNA94} and by Escher $\&$ Simonett \cite{EschSim97}.
In the most general form, the  proof of the existence,
uniqueness of the classical solution and the regularity of the fluid boundary was given by Antontsev,
Gon\c{c}alves and Meirmanov \cite{AntGonMeir02}, \cite{AntGonMeir03}.

{\it Variational} formulation of the Hele-Shaw model was proposed
by Gustafsson \cite{Gust85}, who proved the weak solvability of
the problem (see also \cite{AntMeirYur04}, \cite{GustVas}).

The classical (real-variable) Hele-Shaw model can be reinterpreted
as a mixed boundary value problem for Laplace equation with
respect to unknown parametrization of the boundary and
corresponding Green's function of this problem in the reference
domain. When assuming the presence of a moving obstacle in the flow,
we have to add an additional equation describing this movement.
The aim of our work is to perform an asymptotic analysis
for such a variant of the model and to construct an efficient and robust numerical routine to tackle the problem.

Application of asymptotic methods to approximation of Green's
function goes back to the classical paper by J.~Hadamard
\cite{Had}, where the method of regular perturbation was performed.
Recently, V.~Maz'ya and A.~Movchan obtained a number of asymptotic
formulas for Green's function related to different boundary value
problems for a number of differential operators in
the case of singular perturbations of the domains (see \cite{MazMov09},
\cite{MazMov11}, \cite{MazMovNie13} and references therein).

Those results were used in \cite{MiRoWro1} to model the Hele-Shaw
flow with a fixed circular obstacle.
To construct the computational scheme, we choose Green's formula with the Neumann condition on the external boundary  of doubly connected domain, and with Dirichlet condition on its internal boundary. Then a preliminary transformation $\zeta = \varepsilon/z$ was made, which led to creation of a system of differential equations for the original problem approximation. The existence of the reformulated Hele-Shaw problem follows immediately from the results of the paper Escher-Simonett (1997b).

{The approximate system defined in this way was reduced to the system of first order ODEs, and tackled by a proposed numerical scheme. This approach proved its ability to solve the analyzed Hele-Shaw problem, providing sufficiently good accuracy of computations.}

{However, the scheme itself exhibited some disadvantages:}

\noindent {1) in the case of sink the life-time of the approximate process was very short;}

\noindent {2) in the case of source we get numerical result only on  a bounded  interval of time.}

{The first difficulty could have been expected from the theory of the Hele-Shaw problem, but the second one is a consequence of the numerical scheme instability. In the case of moving obstacle this led to even worse results}.

{In this paper we consider  the Hele-Shaw flow with a rigid inclusion moving in the direction of the flow without rotation. The friction between the limiting planes and the
obstacle is accounted for. To avoid problems appeared in the case of the fixed obstacle,
 we use here the same Maz'ya-Movchan approach, but with different uniform asymptotic formula for Green's function (without making any preliminary transformations, and thus interchanging the role of boundary conditions).
}
%The numerical algorithm has been
%developed improving that proposed in \cite{MiRoWro1}.

The paper is organized as follows. Sec. 2 describes the problem's
geometry  and presents the (real-variable) Hele-Shaw model in a
domain with a moving obstacle. The model is further reduced to the
form containing an unknown parametrization of the boundary of the
fluid domain, an unknown Green's function of the corresponding
mixed boundary value problem for the Laplace operator and unknown
trajectory of the inclusion. Therefore we have to consider a
system of equations consisting of the equation for the free
boundary (the standard Hele-Shaw equation) and the equation of
motion for the obstacle. In Sec. 3 we present uniform asymptotic formula by Maz'ya-Movchan and describe its components.
In Sec. 4 we determine the values of the components of the Maz'ya-Movchan
formula corresponding to the considered model. {The final form of the approximate system of
differential equations is presented in Subsec. 4.4.}
Finally, in Sec. 5  the aforementioned system is implemented in a numerical scheme which illustrates the obtained results. {The numerical scheme  to obtain the solution employs reduction of the system of governing equations to the system of ODEs of the first order, where the velocity of the inclusion is introduced as an additional dependent variable.  In order to solve the dynamic system we utilize the standard ODE solver of Matlab package: ode45. Respective conformal mappings of the boundary curve are performed by means of the Schwartz-Christoffel Toolbox. The derivatives of the mapping along the free boundary are computed by our own subroutines, based on the spline approximation. We show that the used asymptotic expansion for the Green function is effective and the computations based on that approach are stable and robust.}

\section{Problem formulation}

{We consider the slow flow in the Hele-Shaw cell (i.e. in the narrow gap
between to parallel plates of distance $h$). The flow is caused by
a source/sink (situated at the origin $O = (0, 0)$0 of intensity $Q$.
The fluid of the viscosity $\mu$ occupies the bounded doubly connected domain $D_1(t)$ at the time instant $t \geq 0$ that
takes the form $D_1(t) = D(t) \ F$, where $D(t)$ is a bounded simply-connected domain,  and the compact set
$F \subset D(t)$ is a small obstacle embedded within the fluid. The obstacle is moving in the direction of flow
rotation free and with friction coefficient $\kappa$. To avoid technical difficulties, we accept a circular shape of
the obstacle of the radius $\varepsilon$ and of center ${\bf z}_0(t)$ at each instant of time $t$.}

Suppose that our initial geometry satisfies the following conditions\footnote{Note that both constants $c$ and $d$ do not depend on $\delta$.} 
$$
c \leq \min dist\, \{O, D(0)\} \leq \max dist\, \{O, D(0)\} \leq 1,
$$
\begin{equation}
\label{boundary} dist\, \left\{\partial F,
\partial D(0)\right\} = d > \varepsilon,\;\;\; d + 2 \varepsilon > c.
\end{equation}

Following \cite{AntGonMeir03}, the
initial free
boundary $\partial D(0)$ is to satisfy the smoothness
assumptions
\begin{equation}
\label{boundary_smooth}
%\partial\,F\in {\mathcal C}^{2,\alpha},\;\;\;
\partial D(0)\in {\mathcal
C}^{2,\alpha},
\end{equation}
with certain fixed $\alpha, 0 < \alpha < 1$.

Omitting the standard description of the (real-variable) Hele-Shaw model
(see, e.g. \cite{GustVas}, \cite{HohRei}, cf. \cite{MiRoWro1}) we arrive at the following
problem with respect
to unknown parametrization ${\bf w}(s, t)$ of the free boundary $\Gamma(t) = \partial D(t)$
(i.e. the boundary value of the conformal mapping of the unit disc ${\mathbb U}$
onto the fluid domain $D_1(t)$), Green's function ${\mathcal G}({\bf z}; {\bf \zeta}; t)$ of the domain
$D_1(t)$ and the center ${\bf z}_0(t)$ of the obstacle.\footnote{Unknown magnitudes ${\bf w}$, ${\mathcal G}$, ${\bf z}_0$ depend
on time $t$ from a right-sided neighborhood $I$ of $t=0$. In fact, for our problem we need to
determine the value of ${\mathcal G}({\bf z}; {\bf \zeta}; t)$ only at the point ${\bf \zeta} = O$, but we keep the extra variable ${\bf \zeta}$
for computational reasons.}

%Real-variable formulation in terms of diffeomorphisms.
\vspace{2mm} \noindent{\bf Problem (${\bf HS_{move}}$).} Find a
triple $\left\{{\bf w}(s, t) = (w_1(s, t), w_2(s, t)); {\mathcal G}({\bf z}; {\bf \zeta}; t)\right.$; $\left.{\bf z}_0(t) = (z_{0, 1}(t), z_{0, 2}(t))\right\}$,  satisfying the following conditions

(i) ${\bf w}(s, t)\in \Gamma(t)$ for all $(s, t)\in \partial\, {\mathbb
U} \times I$;

(ii) ${\bf w}(\cdot, t) : \partial\, {\mathbb U} \rightarrow \Gamma(t)$
is a ${\mathcal C}^2$-diffeomorphism for each fixed $t\in I$;

(iii) ${\bf w}^{(0)}(s) = {\bf w}(s, 0)$ is a given ${\mathcal
C}^2$-diffeomorphism of the unit circle $\partial\, {\mathbb U}$,
which describes the boundary $\Gamma(0)$ of initial domain
$D_{1}(0)$;

(iv) ${\mathcal G}({\bf z}; {\bf \zeta}; t)$ is Green's function of the operator
$- \triangle$ in the doubly connected domain ${D_{1}(t)}$ with the
homogeneous Neumann condition on $\partial\,F$ and
the homogeneous Dirichlet condition on $\Gamma(t)$;

(v) $\partial_{t}\, {\bf w}(s, t) = - \frac{Q h^2}{12 \mu}\cdot \nabla
{\mathcal G}({\bf w}(s, t); O;  t)$ for all $(s, t)\in
\partial\, {\mathbb U} \times I$;

(vi) ${\bf z}_0^{\prime\prime}  + \frac{\kappa \pi \delta^2}{m}
{\bf z}_0^{\prime} = \frac{Q \delta}{m} \int\limits_{0}^{2 \pi}
{\mathcal G}(z_{0,1} + \delta \cos  \theta,  z_{0,2} + \delta \sin
\theta; O; t) \cdot {\bf n}^{(in)}(\theta) d \theta$, 

(vii) ${\bf z}_0(0)={\bf z}^{(0)}$, ${\bf
z}_0^{\prime}(0)={\bf z}^{(1)}$.

The aim of our study is to get an approximate solution to the problem ${\bf HS_{move}}$.
%Existence of the solution to the above problem can be shown in a way similar to that for existence of the solution for
%the flow in the Hele-Shaw cell with air bubbles in the flow (see, e.g. \cite{EntEti11} and references therein).

\section{Uniform representation of Green's function}

In order to replace the system of equation $(i)-(vii)$ of the Problem (${\bf HS_{move}}$) by the approximate system
we  use one the results  by Maz'ya and Movchan.
For further convenience, we introduce here a small
parameter $\varepsilon$ equal to the radius of the inclusion, and denote by $F_0 = F_0(t)$ the rescaled obstacle $F_0(t) = \{{\bf x}: \frac{1}{\varepsilon} \left({\bf x} - {\bf z}_0(t)\right)\in F\}$. Note that for each $t\in I$ we have $F_0(t) = B(O; 1)$.

For the reader's convenience, we reformulate in our notation the theorem by V.~Maz'ya and A.~Movchan
providing uniform asymptotic approximation of Green's function with the Neumann data on the boundary of the obstacle $F= B({\bf z}_0,\varepsilon)$ and the Dirichlet data on the boundary of the  domain $D$.

\begin{theorem}
\label{MaMov} {\rm{\cite[Thm. 2.1]{MazMovNie13}}} {Let  $D_1$ be a bounded doubly connected domain in ${\mathbb R}^2$ with a smooth boundary, $D_1 = D\setminus F$, where $D$ is a simply connected domain and $F\subset D$ is a compact set (obstacle) with diameter smaller than the distance of $\partial F$ to $\partial D$.}

Green's function ${\mathcal G}_{\varepsilon}({\bf x},{\bf y})$ of the Laplace operator $- \Delta$ with the Neumann zero-data on $\partial F$ and the Dirichlet zero-data on  $\partial D$
has the {following uniform} asymptotic representation
\begin{equation}
\label{MazMov_uniform}
{\mathcal G}_{\varepsilon}({\bf x},{\bf y}) = G({\bf x},{\bf y}) + {\mathcal N}\left(\frac{1}{\varepsilon} ({\bf x} - {\bf z}_0), \frac{1}{\varepsilon} ({\bf y} - {\bf z}_0)\right) + \frac{1}{2 \pi} \log\left|\frac{1}{\varepsilon} ({\bf x} -  {\bf y})\right| +
\end{equation}
$$
+ \varepsilon {\bf{\mathcal D}}\left(\frac{1}{\varepsilon} ({\bf x} - {\bf z}_0)\right) \cdot
\nabla_{\bf x} H({\bf z}_0, {\bf y}) +  \varepsilon {\bf{\mathcal D}}\left(\frac{1}{\varepsilon} ({\bf y} - {\bf z}_0)\right) \cdot
\nabla_{\bf y} H({\bf x}, {\bf z}_0) + {\bf r}_{\varepsilon}({\bf x},{\bf y}),
$$
where
$
 \left|{\bf r}_{\varepsilon}({\bf x},{\bf y})\right| \leq Const \cdot {\varepsilon}^2.
$
\end{theorem}

{In what follows we use this formula in the fluid domain $D_1(t)$ for all $t\in I =[0, T]$,
for which the solution to the Hele-Shaw problem exists. In the case of the flow without obstacle
or with a fixed obstacle we refer for the existence to \cite{AntGonMeir02}. In our case there is
no rigorous proof of the existence, but it can be obtained similarly to that for the flow of bubbles
(see, e.g. \cite{EntEti11} and references therein).}

In our case  we accept in this Theorem the following notation for each instant of time $t\in I$.
$G({\bf x},{\bf y}) = G({\bf x},{\bf y};t)$ is Green's function of the Laplace operator $- \Delta$
for the simply connected domain $D = D(t)$ with zero Dirichlet data on $\partial D(t)$:
\begin{equation}
\label{Green_simply} G({\bf x},{\bf y}) = \frac{1}{2 \pi} \log
\left|\frac{1}{{\bf x} -  {\bf y}}\right| - H({\bf x},{\bf y}),
\end{equation}
with $H$ being the regular part of Green's function, i.e. harmonic function solving the following boundary value problem
\begin{equation}
\label{Green_regular}
\Delta_{\bf x} H({\bf x},{\bf y}) = 0, \;\;\; {\bf x},{\bf y}\in D(t),
\end{equation}
\begin{equation}
\label{Green_regular1}
 H({\bf x},{\bf y}) =  \frac{1}{2 \pi} \log \left|\frac{1}{{\bf x} -  {\bf y}}\right|, \;\;\; {\bf x}\in \partial D(t),{\bf y}\in D(t).
\end{equation}
${\mathcal N}({\bf{\xi}}, {\bf{\eta}})$ is the Neumann function for the exterior of the re-scaled obstacle $F_0 = F_0(t)$:
\begin{equation}
\label{Neumann_regular}
{\mathcal N}({\bf{\xi}}, {\bf{\eta}}) =  \frac{1}{2 \pi} \log \left|{\bf{\xi}} - {\bf{\eta}}\right|^{-1} - h_{\mathcal N}({\bf{\xi}}, {\bf{\eta}}), \;\;\; {\bf{\xi}}, {\bf{\eta}}\in {\mathbb R}^2 \setminus F_0(t),
\end{equation}
where $h_{\mathcal N}({\bf{\xi}}, {\bf{\eta}})$ is the regular part of this function satisfying
\begin{equation}
\label{Neumann_regular1}
\Delta_{\bf{\xi}} h_{\mathcal N}({\bf{\xi}}, {\bf{\eta}}) = 0,\;\;\; {\bf{\xi}}, {\bf{\eta}}\in {\mathbb R}^2 \setminus F_0(t),
\end{equation}
\begin{equation}
\label{Neumann_regular2}
\frac{\partial h_{\mathcal N}({\bf{\xi}}, {\bf{\eta}})}{\partial n_{\bf{\xi}}} = \frac{1}{2 \pi}
\frac{\partial}{\partial n_{\bf{\xi}}} \left(\log \left|{\bf{\xi}} - {\bf{\eta}}\right|^{-1}\right),\;\;\; {\bf{\xi}}\in \partial F_0(t), {\bf{\eta}}\in {\mathbb R}^2 \setminus F_0(t),
\end{equation}
\begin{equation}
\label{Neumann_regular3}
h_{\mathcal N}({\bf{\xi}}, {\bf{\eta}}) \rightarrow 0,\;\;\; |{\bf{\xi}}| \rightarrow \infty, {\bf{\eta}}\in {\mathbb R}^2 \setminus F_0(t).
\end{equation}
The vector-function $ {\bf{\mathcal D}}({\bf{\xi}}) = \left({\mathcal D}_{1}({\bf{\xi}}), {\mathcal D}_{2}({\bf{\xi}})\right)^T$ is the solution of the following boundary value problems in the exterior of the re-scaled obstacle $F_0 = F_0(t)$:
\begin{equation}
\label{Dipole1}
\Delta {\mathcal D}_{j}({\bf{\xi}}) = 0,\;\;\; {\bf{\xi}}\in {\mathbb R}^2 \setminus F_0(t), \; j=1, 2,
\end{equation}
\begin{equation}
\label{Dipole2}
\frac{\partial {\mathcal D}_{j}({\bf{\xi}})}{\partial n} = n_j, \;\;\; {\bf{\xi}}\in \partial F_0(t), \; j=1, 2,
\end{equation}
\begin{equation}
\label{Dipole3}
{\mathcal D}_{j}({\bf{\xi}})  \rightarrow 0,\;\;\; |{\bf{\xi}}| \rightarrow \infty, \; j=1, 2.
\end{equation}
Here $n_j$ are components of the inward unit vector  normal  to the boundary of disc $F_0(t)$.

\section{System of equations for the problem  $HS_{move}$}

\subsection{Green's function ${\mathcal G}_{\varepsilon}$  for the problem  $HS_{move}$}
\label{components}

In this subsection we analyze the components of the representation  (\ref{MazMov_uniform}). Let us first consider the Neumann
function ${\mathcal N}({\bf{\xi}}, {\bf{\eta}})$ having in this case an explicit representation (see, e.g., \cite[p. 68]{Pap}):
\begin{equation}
\label{Neumann-sol}
{\mathcal N}({\bf{\xi}}, {\bf{\eta}}) = - \frac{1}{4\pi} \log |{\bf{\xi}} - {\bf{\eta}}|^2 - \frac{1}{4\pi} \log \left[\frac{(|{\bf{\xi}}|^2 - 1) (|{\bf{\eta}}|^2 - 1) + |{\bf{\xi}} - {\bf{\eta}}|^2}{|{\bf{\xi}}|^2 |{\bf{\eta}}|^2}\right].
\end{equation}
satisfying the conditions (\ref{Neumann_regular}),  (\ref{Neumann_regular1})--(\ref{Neumann_regular3}) and symmetric
${\mathcal N}({\bf{\xi}}, {\bf{\eta}})  ={\mathcal N}({\bf{\eta}}, {\bf{\xi}}).$
Its regular part $h_{\mathcal N}({\bf{\xi}}, {\bf{\eta}})$ is also symmetric and calculated explicitly
yields
\begin{equation}
\label{Neumann-reg1}
h_{\mathcal N}({\bf{\xi}}, {\bf{\eta}}) =  \frac{1}{4\pi} \log \left[\frac{(|{\bf{\xi}}|^2 - 1) (|{\bf{\eta}}|^2 - 1)  + |{\bf{\xi}} - {\bf{\eta}}|^2}{|{\bf{\xi}}|^2 |{\bf{\eta}}|^2}\right].
\end{equation}

Green's function $G({\bf{x}}; {\bf{y}}; t)$ for the interior simply connected domain $D(t)$ {can be represented in the form}
\begin{equation}
\label{Green-int-sol} G({\bf{x}}; {\bf{y}}; t) = - \frac{1}{2\pi}
\log {|g({\bf{x}}, {\bf{y}})|},
\end{equation}
where $g({\bf{x}}, {\bf{y}}) = \left(g_1({\bf{x}}, {\bf{y}}), g_2({\bf{x}}, {\bf{y}})\right): D(t) \rightarrow {\mathbb U}$ {is the conformal mapping of $D(t)$ onto the unit disc ${\mathbb U}$, satisfying} the following normalizing conditions
$g({\bf{x}}, {\bf{y}}){\bigl|_{\bf{x} =  {\bf{y}}}} = 0$, and
$g^{\prime}({\bf{x}}, {\bf{y}}){\bigl|_{\bf{x} =  {\bf{y}}}} > 0$. In our case, $ {\bf{y}} = O$ stands for the  source/sink point (we again note that from computational point of view it is better to keep extra-variable $ {\bf{y}}$ up to the final formula). From the numerical point of view it is customary to start with an arbitrary conformal mapping  $g_{0}({\bf{x}}) : D(t) \rightarrow {\mathbb U}$ and determine the normalized one:
$$
g({\bf{x}}, {\bf{y}}) =
\frac{g_{0}({\bf{x}}) - g_{0}({\bf{y}})}{1 - \overline{g_{0}( {\bf{y}})} g_{0}({\bf{x}})}.
$$

The vector-function $ {\bf{\mathcal D}}({\bf{\xi}})$ can be found by using integral representation of the solution to the exterior Neumann problem for the unit disc (see, e.g., \cite[p. 68]{Pap}). First we note that the inward unit normal vector on the boundary of the unit disc $F_0(t)$ is
\begin{equation}
\label{Normal_in}
{\bf n}^{(in)} = \left(n_1^{(in)}, n_2^{(in)}\right) = - (\cos \varphi, \sin \varphi),
\end{equation}
where $\varphi$ is the angular coordinate of polar system on the unit circle $\partial F_0(t)$. Then the solutions to the problems (\ref{Dipole1})--(\ref{Dipole3}) ($j = 1, 2$)
are represented in the form
\begin{equation}
\label{Dipole_sol1}
{\mathcal D}_{j}({\bf{\xi}}) = \frac{1}{2 \pi} \int\limits_{0}^{2 \pi} \log \left(\frac{1}{r^2} + 1 - \frac{2}{r} \cos (\theta - \varphi)\right) n_j^{(in)} d \varphi,\;\;\; j=1, 2,
\end{equation}
where ${\bf{\xi}} = (r \cos \theta, r \sin \theta), r > 1$. It is easy to see that the above functions
satisfy all conditions (\ref{Dipole1})--(\ref{Dipole3}). We calculate (\ref{Dipole_sol1})  using formula \cite[(4.397.6)]{GrRyzh}:
\begin{equation}
\label{Dipole_repr}
{\mathcal D}_{1}({\bf{\xi}}) = \frac{1}{2} \frac{\xi_1}{\xi_1^2 + \xi_2^2},\;\;\;
{\mathcal D}_{2}({\bf{\xi}}) = \frac{1}{2} \frac{\xi_2}{\xi_1^2 + \xi_2^2}.
\end{equation}

\subsection{Derivatives of Green's function}
\label{DER}

Here we calculate derivatives of Green's function which are used in equation (v) of the Problem (${\bf HS_{move}}$). We start with Green's function $G({\bf x}; {\bf y}) = G({\bf x}; {\bf y}; t)$. By applying representation (\ref{Green-int-sol}) we have ($j = 1, 2$)
\begin{equation}
\label{Green_der1}
\partial_{x_j} G({\bf x}; {\bf y}) = - \frac{1}{2 \pi} \frac{g_1({\bf x}; {\bf y}) \partial_{x_j} g_1({\bf x}; {\bf y}) + g_2({\bf x}; {\bf y}) \partial_{x_j} g_2({\bf x}; {\bf y})}{g_1^2({\bf x}; {\bf y}) + g_2^2({\bf x}; {\bf y})}.
\end{equation}
Substituting ${\bf x} = (w_1(s, t), w_2(s, t))$, ${\bf y} = (0, 0)$ and taking into account the properties of the function $g({\bf x}; {\bf y})$ we finally obtain
\begin{equation}
\label{Green_der2}
\partial_{x_j} G(w_1(s, t), w_2(s, t); 0, 0) =
\end{equation}
$$
= - \frac{1}{2\pi} \left(g_1(w_1(s, t), w_2(s, t); 0, 0)
\partial_{w_j} g_1(w_1(s, t), w_2(s, t); 0, 0) +\right.
$$
$$
\left.+ g_2(w_1(s, t), w_2(s, t); 0, 0) \partial_{w_j} g_2(w_1(s, t), w_2(s, t); 0, 0)\right).
$$

The Neumann function ${\mathcal N}({\bf{\xi}}; {\bf{\eta}})$ depends on the ``scaled'' variables
\begin{equation}
\label{scaled}
{\bf{\xi}} = \frac{1}{\varepsilon} \left({\bf x} - {\bf z}_0(t)\right), \;\;\;
{\bf{\eta}} = \frac{1}{\varepsilon} \left({\bf y} - {\bf z}_0(t)\right).
\end{equation}
Hence
$$
\partial_{x_j} {\mathcal N}({\bf{\xi}}; {\bf{\eta}}) =
\partial_{\xi_j} {\mathcal N}({\bf{\xi}}; {\bf{\eta}}) \frac{\partial \xi_j}{\partial x_j}=
 \frac{1}{\varepsilon} \partial_{\xi_j} {\mathcal N}({\bf{\xi}}; {\bf{\eta}}).
$$
Using explicit representation of the Neumann function (\ref{Neumann-sol}) we get the following value of the derivatives $(j = 1, 2)$
\begin{equation}
\label{Neumann_der1}
\partial_{x_j} {\mathcal N}({\bf{\xi}}; {\bf{\eta}}) = - \frac{1}{2 \pi \varepsilon}
\frac{\xi_j - \eta_j}{(\xi_1 - \eta_1)^2 + (\xi_2 - \eta_2)^2} + \frac{1}{2 \pi \varepsilon}  \frac{\xi_j}{\xi_1^2 + \xi_2^2} -
\end{equation}
$$
 - \frac{1}{2 \pi \varepsilon}\left[
\frac{\xi_j(\eta_1^2 + \eta_2^2) - \eta_j}{(\xi_1^2 + \xi_2^2 - 1)(\eta_1^2 + \eta_2^2 - 1) +
(\xi_1 - \eta_1)^2 + (\xi_2 - \eta_2)^2}\right].
$$
Now substitute ${\xi}_j = \frac{1}{\varepsilon} \left(w_j(s,t) -  z_{0,j}(t)\right)$, $\eta_j =
- \frac{1}{\varepsilon} z_{0,j}(t)$, $j = 1, 2$ (in order to simplify representation we omit internal variables $s$ and $t$ in the right hand-side of this relation) and 
 calculate derivatives of two terms of Maz'ya-Movchan asymptotic formula (see (3.1)) at ${\bf x} = (w_1, w_2)$ with ${\bf y} = (0, 0)$
$$
\frac{\partial}{\partial x_j} \left({\mathcal N}(\xi, \eta) + \frac{1}{4\pi} \log\, |\frac{1}{\varepsilon} ({\bf x} - {\bf y})|^2\right).
$$
These derivatives (denoted  $K_j$) are are equal
\begin{equation}
\label{K_j}
K_j = \frac{\partial}{\partial x_j} \left({\mathcal N}(\xi, \eta) + \frac{1}{4\pi} \log\, |\frac{1}{\varepsilon} ({\bf x} - {\bf y})|^2\right) =
\end{equation}
$$
 = - \frac{1}{2\pi} \left\{\frac{(w_j - z_{0,j})[z_{0,1}^2 + z_{0,2}^2 - \varepsilon^2] + \varepsilon^2 w_j}{[(w_1 - z_{0,1})^2 + (w_2 - z_{0,2})^2 - \varepsilon^2][z_{0,1}^2 + z_{0,2}^2 - \varepsilon^2] + \varepsilon^2 (w_1^2 + w_2^2)} - \right.
$$
$$
\left. - \frac{w_j - z_{0,j}}{(w_1 - z_{0,1})^2 + (w_2 - z_{0,2})^2} \right\}.
$$
Thus for $\varepsilon = 0$ we have $K_j = 0$.

For regular part $H({\bf x}; {\bf y})$ of Green's function $G({\bf x}; {\bf y})$ we have the representation (\ref{Green_simply}), i.e.
$$
H({\bf x}; {\bf y}) = \frac{1}{2\pi} \log\left|\frac{g({\bf x}; {\bf y})}{{\bf x} - {\bf y}}\right|.
$$
Therefore
\begin{equation}
\label{Gr_reg_der_x1}
\partial_{x_j} H({\bf w}; O) = + \frac{1}{2 \pi} \frac{g_1({\bf z}_0; O) \partial_{z_{0,j}} g_1({\bf z}_0; O)
+ g_2({\bf z}_0; O) \partial_{z_{0,j}} g_2({\bf z}_0;
O)}{g_1^2({\bf z}_0; O) + g_2^2({\bf z}_0; ))} -
\end{equation}
$$
- \frac{1}{2 \pi} \frac{z_{0,j}}{z_{0,1}^2 + z_{0,2}^2} =
$$
$$
 = \frac{1}{2 \pi} \left( g_1({\bf w}; O) \partial_{w_j} g_1({\bf w}; O) + g_2({\bf w}; O) \partial_{w_j} g_2({\bf w}; O)
 - \frac{w_{j}}{w_1^2 + w_2^2}\right).
$$

Analogously,
\begin{equation}
\label{Gr_reg_der_y1}
\partial_{y_j} H({\bf x}; {\bf z}_0) = + \frac{1}{2 \pi}
\frac{g_1({\bf x}; {\bf z}_0)
\partial_{y_j} g_1({\bf x}; {\bf z}_0) + g_2({\bf
x}; {\bf z}_0) \partial_{y_j} g_2({\bf x}; {\bf z}_0)}{g_1^2({\bf
x}; {\bf z}_0) + g_2^2({\bf x}; {\bf z}_0)} +
\end{equation}
$$
+ \frac{1}{2 \pi} \frac{x_j - z_{0,j}}{(x_1 - z_{0,1})^2 + (x_2 -
z_{0,2})^2}.
$$
In this case the right hand-side of the last relation does depend on ${\bf x}$.

Now we have to calculate the derivatives with respect to ${\bf x}$ of the following expression
$$
J_1({\bf x}, {\bf y}) := J_1 = \varepsilon  {\bf{\mathcal D}}\left(\frac{1}{\varepsilon} ({\bf x} - {\bf z}_0)\right) \cdot \nabla_{\bf x} H({\bf z}_0; {\bf y}) =
$$
$$
= \varepsilon {\mathcal D}_{1}\left(\frac{1}{\varepsilon} ({\bf x} - {\bf z}_0)\right) \partial_{x_1} H({\bf z}_0; {\bf y}) + \varepsilon {\mathcal D}_{2}\left(\frac{1}{\varepsilon} ({\bf x} - {\bf z}_0)\right) \partial_{x_2} H({\bf z}_0; {\bf y}),
$$
where only first multiplier in each summand depends on ${\bf x}$. Derivatives of ${\mathcal D}_k$ in $x_j$ is connected with that in $\xi_j$
$$
\partial_{x_j} \left(\varepsilon  {\mathcal D}_k\left(\frac{1}{\varepsilon} ({\bf x} - {\bf z}_0)\right)\right) = \left({\mathcal D}_k\right)^{\prime}_{\xi_j}\left(\frac{1}{\varepsilon} ({\bf x} - {\bf z}_0)\right),\; j, k = 1, 2.
$$
By the direct calculation  we have
\begin{equation}
\label{Dipole_derJ1_1}
\partial_{x_1} J_1({\bf w}, O) = \frac{\varepsilon^2}{2} \left(\frac{(w_2 - z_{0,2})^2 - (w_1 - z_{0,1})^2}{((w_1 - z_{0,1})^2 + (w_2 - z_{0,2})^2)^2} \cdot \partial_{x_1} H({\bf z}_0; O) - \right.
\end{equation}
$$
\left.- \frac{2 (w_1 - z_{0,1}) (w_2 - z_{0,2})}{((w_1 - z_{0,1})^2 + (w_2 - z_{0,2})^2)^2} \cdot \partial_{x_2} H({\bf z}_0; O)\right),
$$
and
\begin{equation}
\label{Dipole_derJ1_2}
\partial_{x_2} J_1({\bf w}, O) = \frac{\varepsilon^2}{2} \left(- \frac{2 (w_1 - z_{0,1}) (w_2 - z_{0,2})}{((w_1 - z_{0,1})^2 + (w_2 - z_{0,2})^2)^2} \cdot \partial_{x_1} H({\bf z}_0; O) - \right.
\end{equation}
$$
\left. - \frac{(w_2 - z_{0,2})^2 - (w_1 - z_{0,1})^2}{((w_1 - z_{0,1})^2 + (w_2 - z_{0,2})^2)^2} \cdot \partial_{x_2} H({\bf z}_0; O)\right),
$$
where the derivatives $\partial_{x_j} H({\bf z}_0; {\bf y})$ are presented in (\ref{Gr_reg_der_x1}).

At last we have
$$
J_2({\bf x}, {\bf y}) = J_2 = \varepsilon  {\bf{\mathcal D}}\left(\frac{1}{\varepsilon} ({\bf y} - {\bf z}_0)\right) \cdot \nabla_{\bf y} H({\bf x}; {\bf z}_0) =
$$
$$
\varepsilon {\mathcal D}_{1}\left(\frac{1}{\varepsilon} ({\bf y} - {\bf z}_0)\right) F_1({\bf x}; {\bf z}_0) + \varepsilon {\mathcal D}_{2}\left(\frac{1}{\varepsilon} ({\bf y} - {\bf z}_0)\right) F_2({\bf x}; {\bf z}_0).
$$
Here, only second multiplier in each summand depends on ${\bf x}$, and
$$
F_i = \partial_{y_i} H({\bf x}; {\bf z}_0) =  \frac{1}{2\pi}
\partial_{y_i} \log \left|\frac{g({\bf x}; {\bf y})}{{\bf x} -
{\bf y}}\right|_{{\bf y} = {\bf z}_0}, \;\;\; i = 1, 2.
$$

Hence
$$
\partial_{x_j} J_2 = \frac{\varepsilon}{2} \frac{\eta_1}{\eta_1^2 + \eta_2^2} \partial_{x_j} F_1  + \frac{\varepsilon}{2} \frac{\eta_2}{\eta_1^2 + \eta_2^2} \partial_{x_j} F_2, \quad j = 1, 2.
$$

Therefore, since $\eta_i = - \frac{1}{\varepsilon} z_{0,i},\, i = 1, 2,$ and using (\ref{Dipole_repr}) we have
\begin{equation}
\label{Dipole_derJ2}
\partial_{x_j} J_2({\bf w}, O) =  - \frac{\varepsilon^2}{2} \frac{z_{0,1}}{z_{0,1}^2 + z_{0,1}^2} \cdot \partial_{x_j} F_1({\bf w}; {\bf z}_0) -
 \frac{\varepsilon^2}{2} \frac{z_{0,2}}{z_{0,1}^2 + z_{0,1}^2} \cdot \partial_{x_j} F_2({\bf w}; {\bf z}_0).
\end{equation}

\subsection{Integrals of Green's function}
\label{INT}

In this subsection we calculate integrals from the right hand-side of the equation (vi) in the representation of {\bf Problem (${\bf HS_{move}}$)}:
\begin{equation}
\label{Green-int1} P_j = \frac{Q \varepsilon}{m}
\int\limits_{0}^{2 \pi} {\mathcal G}({\bf x}; {\bf y}; t)
{n}^{(in)}_j(\theta) d \theta,\;\;\; j=1, 2,
\end{equation}
were ${\bf x} = (z_{0,1} + \varepsilon \cos \theta, z_{0,2} + \varepsilon \sin \theta)$, ${\bf y} = (0, 0)$.

In our calculations we use
components of formula (\ref{MazMov_uniform}) and their representations obtained in Subsec. \ref{components}.
First we calculate the integral
\begin{equation}
\label{Green-int11}
I_{1,1} := \int\limits_{0}^{2\pi} G({\bf x}; {\bf y}; t) \cos \theta d \theta \left|_{{\bf y} = O}\right.
\end{equation}
employing representation (\ref{Green-int-sol}):
$$
I_{1,1} = - \frac{1}{2 \pi} \int\limits_{0}^{2\pi} \log
\left|{g({\bf x}; O)}\right|  \cos \theta d \theta =
 - \frac{1}{2 \pi} {\mathrm{Re}} \int\limits_{0}^{2\pi} \log {g({\bf x}; O)} \cos \theta d \theta.
$$
Note that for each fixed $t\in I$ the function $\log {g({\bf x}; O)}$ is an analytic function with respect to variable ${\bf x} = (x_1, x_2)$ in the disc $B({\bf z}_0, \varepsilon)$. Hence using Taylor expansion of $\log {g({\bf x}; O)}$ at ${\bf x} = {\bf z}_0$ we have
\begin{equation}
\label{Green-int11_f} I_{1,1} = - \frac{\varepsilon}{2}
{\mathrm{Re}\, c_1} + O(\varepsilon^3) = - \frac{\varepsilon}{2}
{\mathrm{Re}\, \frac{g^{\prime}({\bf z}_0, O)}{g({\bf z}_0, O)}} +
O(\varepsilon^3).
\end{equation}

Similar calculations can be performed for the integral
\begin{equation}
\label{Green-int11_tilde}
\tilde{I}_{1,1} := \int\limits_{0}^{2\pi} G({\bf x}; {\bf y}; t) \sin \theta d \theta \left|_{{\bf y} = O}\right. = -
\frac{\varepsilon}{2} {\mathrm{Re}\, i c_1} + O(\varepsilon^3) = -
\frac{\varepsilon}{2} {\mathrm{Re}\, \frac{i g^{\prime}({\bf z}_0,
O)}{g({\bf z}_0, O)}} + O(\varepsilon^3).
\end{equation}

In the integral
\begin{equation}
\label{Green-int12}
I_{1,2} := \int\limits_{0}^{2\pi} \left[{\mathcal N}\left(\frac{1}{\varepsilon}({\bf x} - {\bf z}_0, {\bf y} - {\bf z}_0)\right) +
\frac{1}{2 \pi} \log\left|\frac{1}{\varepsilon}({\bf x} - {\bf y})\right|\right] \cos \theta d \theta \left|_{{\bf y} = O}\right.,
\end{equation}
we use representation (\ref{Neumann-sol})
\begin{equation}
\label{Green-int12}
I_{1,2} := \int\limits_{0}^{2\pi} \left[{\mathcal N}\left(\frac{1}{\varepsilon}({\bf x} - {\bf z}_0, - {\bf z}_0)\right) +
\frac{1}{2 \pi} \log\left|\frac{{\bf x}}{\varepsilon}\right|\right] \cos \theta d \theta =
\end{equation}
$$
 = - \frac{1}{4 \pi}  \int\limits_{0}^{2\pi} \log\left\{\left(\frac{|{\bf x} - {\bf z}_0|^2}{\varepsilon^2} - 1\right)
\left(\frac{|{\bf z}_0|^2}{\varepsilon^2} - 1\right) + \frac{|{\bf x}|^2}{\varepsilon^2}\right\} \cos \theta d \theta +
$$
$$
+ \frac{1}{4 \pi}  \int\limits_{0}^{2\pi} \log \left(\frac{|{\bf x} - {\bf z}_0|^2 |{\bf z}_0|^2}{\varepsilon^4}\right) \cos \theta d \theta.
$$
Since $|{\bf x} - {\bf z}_0| = \varepsilon$ and $\int_{0}^{2\pi}
 C \cdot \cos \theta d \theta = 0$, then we have
$$
I_{1,2} = - \frac{1}{2 \pi} \int\limits_{0}^{2\pi} \log |{\bf x}|  \cos \theta d \theta = - \frac{1}{2 \pi} {\mathrm{Re}} \int\limits_{0}^{2\pi} \log {\bf x}  \cos \theta d \theta.
$$

The function $\log {\bf x}$ is an analytic function with respect to variable ${\bf x} = (x_1, x_2)$ in the disc $B({\bf z}_0, \varepsilon)$. Hence 
\begin{equation}
\label{Green-int12_f}
I_{1,2} = - \frac{\varepsilon}{2} {\mathrm{Re}}\, \frac{1}{{\bf z}_0} + O(\varepsilon^3).
\end{equation}
Analogously for the integral $\tilde{I}_{1,2}$ we have 
\begin{equation}
\label{Green-int12}
\tilde{I}_{1,2} := \int\limits_{0}^{2\pi} \left[{\mathcal N}\left(\frac{1}{\varepsilon}({\bf x} - {\bf z}_0, {\bf y} - {\bf z}_0)\right) +
\frac{1}{2 \pi} \log\left|\frac{1}{\varepsilon}({\bf x} - {\bf y})\right|\right] \sin \theta d \theta \left|_{{\bf y} = O}\right.=
\end{equation}
$$
= - \frac{1}{2 \pi} \int\limits_{0}^{2\pi} \log |{\bf x}|  \sin \theta d \theta = - \frac{1}{2 \pi} {\mathrm{Re}} \int\limits_{0}^{2\pi} \log {\bf x}  \sin \theta d \theta = - \frac{\varepsilon}{2} {\mathrm{Re}}\, \frac{i}{{\bf z}_0} + O(\varepsilon^3).
$$

Next we calculate
\begin{equation}
\label{Green-int13}
{I}_{1,3} := \int\limits_{0}^{2\pi} \varepsilon {\bf{\mathcal D}}\left(\frac{1}{\varepsilon}({\bf x} - {\bf z}_0)\right) \cdot \nabla_{\bf x} H({\bf z}_0; {\bf y})  \cos \theta d \theta.
\end{equation}

Taking into account exact values (\ref{Dipole_repr}) of the functions ${\mathcal D}_1({\bf{\xi}})$, ${\mathcal D}_2({\bf{\xi}})$ and parametrization of the boundary of $B(0; 1)$ ($\xi_1 = \cos\, \theta, \xi_2 = \sin\, \theta$) we obtain
$$
{I}_{1,3} = \frac{\varepsilon}{2} \partial_{x_1} H({\bf z}_0; {\bf y}) \int\limits_{0}^{2\pi}   \cos^2 \theta d \theta +
 \frac{\varepsilon}{2} \partial_{x_2} H({\bf z}_0; {\bf y})  \int\limits_{0}^{2\pi}  \sin \theta \cos \theta d \theta = \frac{\pi \varepsilon}{2} \partial_{x_1} H({\bf z}_0; {\bf y}).
$$
Finally, using (\ref{Gr_reg_der_x1})
\begin{equation}
\label{Green-int13_f} {I}_{1,3} = \frac{\varepsilon}{4}
\left(\frac{g_1({\bf z}_0; O) \partial_{x_1} g_1({\bf z}_0; O) +
g_2({\bf z}_0; O) \partial_{x_1} g_2({\bf z}_0; O)}{g_1^2({\bf
z}_0; O) + g_2^2({\bf z}_0; O)} - \frac{z_{0,1}}{z_{0,1}^2 +
z_{0,2}^2}\right).
\end{equation}

Similar calculations lead
\begin{equation}
\label{Green-int13tilde}
\tilde{I}_{1,3} := \int\limits_{0}^{2\pi} \varepsilon {\bf{\mathcal D}}\left(\frac{1}{\varepsilon}({\bf x} - {\bf z}_0)\right) \cdot \nabla_{\bf x} H({\bf z}_0; {\bf y})  \sin \theta d \theta = \frac{\pi \varepsilon}{2} \partial_{x_2} H({\bf z}_0; {\bf y}),
\end{equation}
\begin{equation}
\label{Green-int13_tilde_f} \tilde{I}_{1,3} =
\frac{\varepsilon}{4} \left(\frac{g_1({\bf z}_0; O) \partial_{x_2}
g_1({\bf z}_0; O) + g_2({\bf z}_0; O) \partial_{x_2} g_2({\bf
z}_0; O)}{g_1^2({\bf z}_0; O) + g_2^2({\bf z}_0; O)} -
\frac{z_{0,2}}{z_{0,1}^2 + z_{0,2}^2}\right).
\end{equation}

The last integral from  (\ref{Green-int1}) is calculated by using
(\ref{Dipole_repr})
\begin{equation}
\label{Green-int14}
{I}_{1,4} := \int\limits_{0}^{2\pi} \varepsilon {\bf{\mathcal D}}\left(\frac{1}{\varepsilon}({\bf y} - {\bf z}_0)\right) \cdot \nabla_{\bf y} H({\bf x}; {\bf z}_0)  \cos \theta d \theta =
\end{equation}
$$
= - \frac{\varepsilon^2}{2 (z_{0,1}^2 + z_{0,2}^2)} \int\limits_{0}^{2\pi} \left[z_{0,1} \partial_{y_1} H({\bf x}; {\bf z}_0) + z_{0,2} \partial_{y_2} H({\bf x}; {\bf z}_0)\right] \cos \theta d \theta.
$$

Since $\log \frac{g({\bf x}; {\bf z}_0)}{{\bf x} - {\bf z}_0}$ is an analytic function with respect to variable ${\bf x} = (x_1, x_2)$ in the disc $B({\bf z}_0, \varepsilon)$ then
$$
- \frac{Q \varepsilon}{m} {I}_{1,4} = O(\varepsilon^4).
$$
Similar result we have for the integral
\begin{equation}
\label{Green-int14_f_tilde}
\tilde{I}_{1,4} := \int\limits_{0}^{2\pi} \varepsilon {\bf{\mathcal D}}\left(\frac{1}{\varepsilon}({\bf y} - {\bf z}_0)\right) \cdot \nabla_{\bf y} H({\bf x}; {\bf z}_0)  \sin \theta d \theta,
\end{equation}
$$
- \frac{Q \varepsilon}{m} \tilde{I}_{1,4} = O(\varepsilon^4).
$$

Note that
$$
{\mathrm{Re}}\, \frac{g^{\prime}({\bf z}_0, O)}{g({\bf z}_0, O)} =
\frac{g_1({\bf z}_0; O) \partial_{x_1} g_1({\bf z}_0; O) +
g_2({\bf z}_0; O) \partial_{x_1} g_2({\bf z}_0; O)}{g_1^2({\bf
z}_0; O) + g_2^2({\bf z}_0; O)},
$$
$$
{\mathrm{Re}}\, i \frac{g^{\prime}({\bf z}_0, O)}{g({\bf z}_0, O)}
= \frac{g_1({\bf z}_0; O) \partial_{x_2} g_1({\bf z}_0; O) +
g_2({\bf z}_0; O) \partial_{x_2} g_2({\bf z}_0; O)}{g_1^2({\bf
z}_0; O) + g_2^2({\bf z}_0; O)},
$$
$$
{\mathrm{Re}}\, \frac{1}{{\bf z}_0} = \frac{z_{0,1}}{z_{0,1}^2 +
z_{0,2}^2},\;\;\; {\mathrm{Re}}\, \frac{i}{{\bf z}_0} =
\frac{z_{0,2}}{z_{0,1}^2 + z_{0,2}^2}.
$$

Therefore, combining all above calculations we have
\begin{equation}
\label{int_all} - \frac{Q \varepsilon}{m} \left[I_{1,1} + I_{1,2} +
I_{1,3} + I_{1,4}\right] = \frac{Q \varepsilon^2}{4 m} I +
O(\varepsilon^4),
\end{equation}
\begin{equation}
\label{int_all_tilde} - \frac{Q \varepsilon}{m}
\left[\tilde{I}_{1,1} + \tilde{I}_{1,2} + \tilde{I}_{1,3} +
\tilde{I}_{1,4}\right] = \frac{Q \varepsilon^2}{4 m} \tilde{I} +
O(\varepsilon^4),
\end{equation}
$$
I = \frac{g_1({\bf z}_0; O) \partial_{x_1} g_1({\bf z}_0; O) +
g_2({\bf z}_0; O) \partial_{x_1} g_2({\bf z}_0; O)}{g_1^2({\bf
z}_0; O) + g_2^2({\bf z}_0; O)} + 3 \frac{z_{0,1}}{z_{0,1}^2 +
z_{0,2}^2},
$$
$$
\tilde{I} = \frac{g_1({\bf z}_0; O) \partial_{x_2} g_1({\bf z}_0;
O) + g_2({\bf z}_0; O) \partial_{x_2} g_2({\bf z}_0;
O)}{g_1^2({\bf z}_0; O) + g_2^2({\bf z}_0; O)} + 3
\frac{z_{0,2}}{z_{0,1}^2 + z_{0,2}^2}.
$$

\subsection{Final system of differential equations}
\label{FIN}

It follows from the potential theory (see, e.g. \cite[Ch. 8]{GilTru01}, cf. \cite[Lemma 2.3]{MazMovNie13}), that for any compact subset
$\Omega, \overline{\Omega} \subset D_1(t)$,
$$
\left|\left(r_{\varepsilon}({\bf x}, {\bf y})\right)^{\prime}_{x_j}\right| \leq \varepsilon^3, \quad j=1,2, \quad {\bf x}, {\bf y}\in  \overline{\Omega}.
$$

Thus, the {\bf Problem (${\bf HS_{move}}$)} can be asymptotically approximated by the following system
\begin{equation}
\label{Final_system1}
\partial_t w_1 = - \frac{Q h^2}{12 \mu} \left(\partial_{x_1} G({\bf w}; O) +
 K_1 +
\partial_{x_1} J_1 ({\bf w}; O) + \partial_{x_1} J_2 ({\bf w}; O)\right),
\end{equation}
\begin{equation}
\label{Final_system2}
\partial_t w_2 = - \frac{Q h^2}{12 \mu} \left(\partial_{x_2} G({\bf w}; O) +
K_2 +
\partial_{x_2} J_1 ({\bf w}; O) + \partial_{x_2} J_2 ({\bf w}; O)\right),
\end{equation}
\begin{equation}
\label{Final_system3}
 z_{0,1}^{\prime\prime} + \frac{\kappa \pi \varepsilon^2}{m}  z_{0,1}^{\prime} = \frac{Q \varepsilon^2}{4
 m} I,
\end{equation}
\begin{equation}
\label{Final_system4}
 z_{0,2}^{\prime\prime} + \frac{\kappa \pi \varepsilon^2}{m}  z_{0,2}^{\prime} = \frac{Q \varepsilon^2}{4 m}
\tilde{I},
\end{equation}
with initial conditions ${\bf z}_0(0)={\bf z}^{(0)}$, ${\bf
z}_0^{\prime}(0)={\bf z}^{(1)}$. Here ${\bf w} = (w_1(s, t), w_2(s, t))$ is an unknown parametrization of the external boundary
$\partial D(t)$, ${\bf z}_0 = (z_{0,1}(t),z_{0,2}(t))$ is an
unknown position of the center of the moving obstacle and $K_j$,
$\partial_{x_j} J_k$, $I$, $\tilde{I}$ are defined in (\ref{K_j}),
(\ref{Dipole_derJ1_1}), (\ref{Dipole_derJ1_2}),
(\ref{Dipole_derJ2}), (\ref{int_all}), (\ref{int_all_tilde}).

\section{Numerical examples and discussions}
In this section we provide only a short illustration of efficiency of the proposed methods for applications.
%The computations were done for 70 nodes spaced at the uniform angular distances.
%Comprehensive description of the numerical procedure and
%complete numerical results can be found in \cite{Arxiv}.

{The numerical scheme  to obtain the solution employs reduction of the system of governing equations \eqref{Final_system1}-\eqref{Final_system2} to the system of ODEs of the first order, where the velocity of the inclusion is introduced as an additional dependent variable.  In order to solve the dynamic system of the first order derived in this way we utilize the standard ODE solver of Matlab package: ode45. It is based on  an explicit Runge-Kutta formula. Respective conformal mappings of the boundary curve are performed by means of the Schwartz-Christoffel Toolbox \cite{Dri96}, \cite{Dri05}. The derivatives of the mapping along the free boundary are computed by our own subroutines, based on the spline approximation.}

\begin{figure}[h!]
\centering
    \includegraphics [scale=0.43]{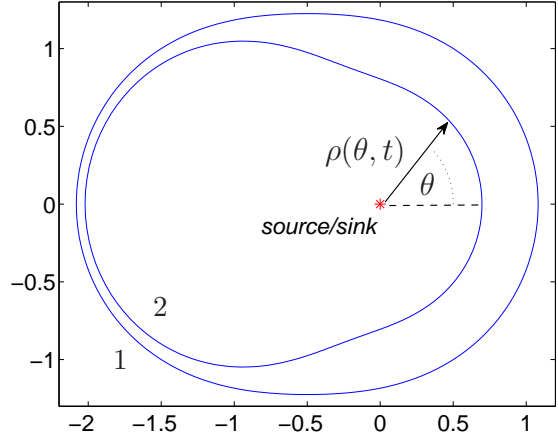}
    \put(-110,114){$\rho (\theta,t)$}
    \put(-74,101){$\theta$}
    %\put(-150,84){$source/sink$}
    \put(-190,35){\small 1}
     \put(-175,55){\small 2}

    \caption{Evolution of the computational domain in time - the limiting curves.  Both cases, contraction from the curve 1 to the curve 2 in case of the sink and expansion
    over the same time (from the curve 2 to the curve 1 in case of the source), are considered.
    Vector $\rho (\theta,t)$ defines the boundary curve.}

    \begin{picture}(0,0)(70,100)
    \label{domain_shape}
    \end{picture}
\end{figure}

\begin{figure}[h!]
\centering
    \includegraphics [scale=0.43]{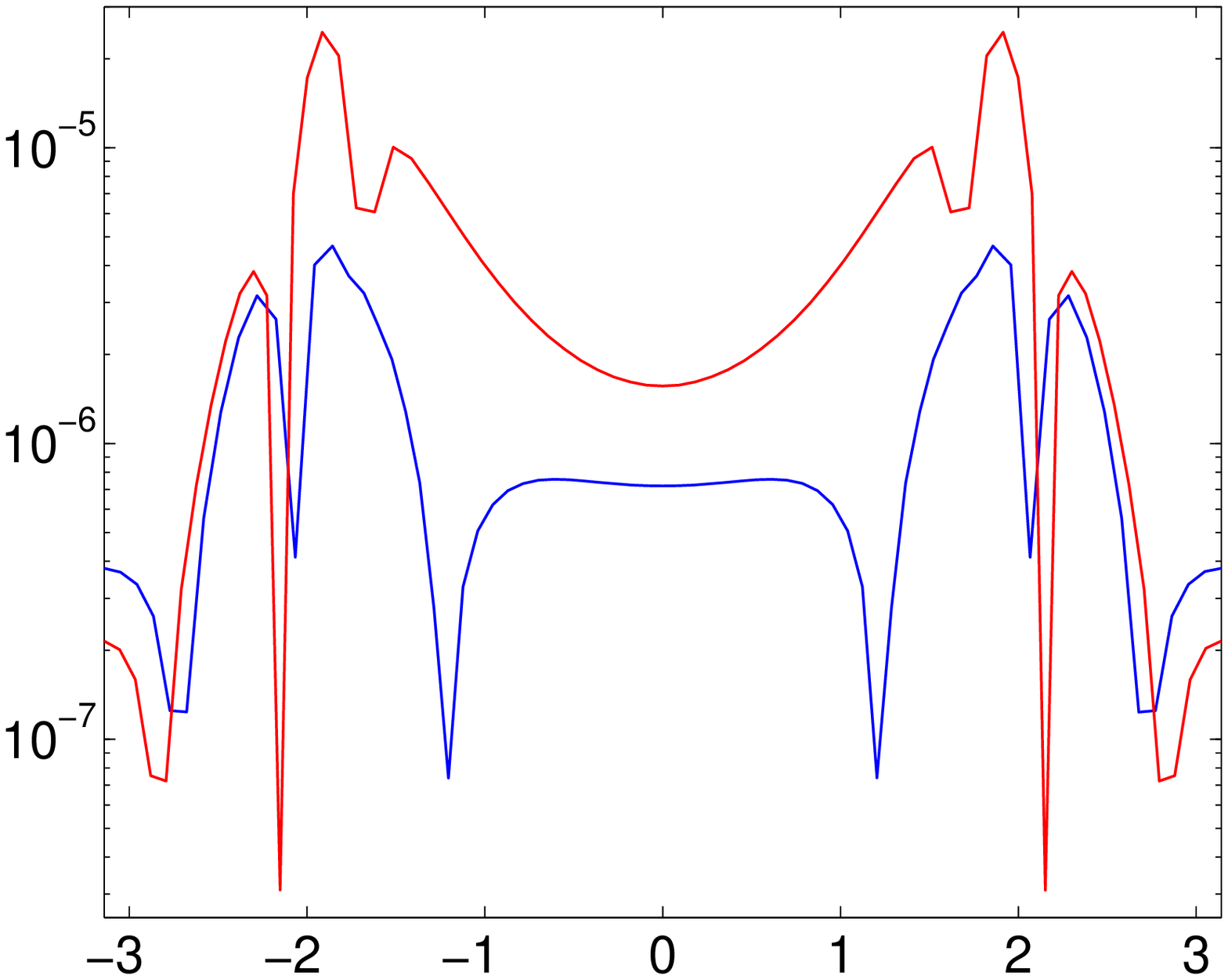}
    \put(-120,0){$\theta$}
    \put(-245,100){$\delta \rho$}
    \caption{Relative error of the radius vector $\rho(t_{max},\theta)$.  Blue line corresponds the source case, while the red one refers to the sink configuration.}
    \begin{picture}(0,0)(70,100)
    \label{rho_error}
    \end{picture}
\end{figure}

{To investigate the accuracy of the proposed numerical scheme we use the classical benchmark by Polubarinova-Kochina \cite[p. 29]{GustVas}, which describes the fluid domain induced by a source or a sink without inclusion ($\varepsilon =0$). Evolution of the free boundary in the considered case is illustrated in Fig.\ref{domain_shape}. We analyze three different densities of the spatial meshing, described by the number of the nodes, $N$, distributed at uniform angular distances: $N=35$, $N=70$, $N=120$. Moreover, both the fluid source and fluid sink variants are considered. In the first case the free boundary evolves from the internal to external shape (see  Fig.\ref{domain_shape}). In the second one it moves in reverse direction. The results of computations illustrated by the relative error of the radius vector defining the free boundary, $\delta \rho$, are shown in
Fig.\ref{rho_error}.}

It shows that the solution accuracy  is of one order of magnitude better for the expansion ($\delta \rho =10^{-6}$) than that for the contraction ($\delta \rho =10^{-5}$) of the domain. Thus, the proposed algorithm is capable of tackling both cases with a satisfactory solution accuracy.

\begin{figure}[h!]
\centering
    \includegraphics [scale=0.45]{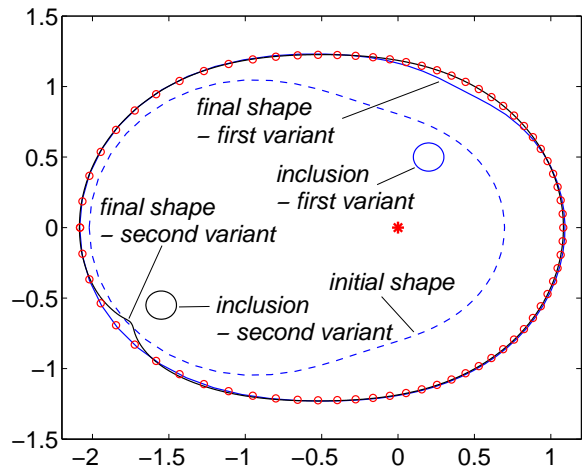}
    %\put(-120,0){$\theta$}
    %\put(-245,100){$\delta \rho$}
    \caption{Graphical illustration of the problem with an immobile inclusion. Two locations of the obstacle are considered.
    The curve with markers corresponds to the results for undisturbed flow (without inclusion).}
    \begin{picture}(0,0)(70,100)
    \label{imm_incl_shapes}
    \end{picture}
\end{figure}

In the next step we investigate to what degree the presence of an immobile inclusion inside the domain affects the fluid flow. Now, we restrict ourselves only to the case of fluid source and consider a circular inclusion of the radius $\varepsilon= 0.2$ inside the domain encircled by the internal curve 2 from the previous benchmark. We retain the same source intensity and time interval assuming zero initial conditions (${\bf z}^{(0)}={\bf z}^{(1)}=0$) in the absence of any forces in the right-hand sides of \eqref{Final_system3} - \eqref{Final_system4}. Two various locations of the inclusion are considered: ${\bf z}_0=0.2+0.5i$ and ${\bf z}_0=-1.55-0.55i$. The graphical illustration of the problem is shown in Fig. \ref{imm_incl_shapes}, where the final shapes for the free boundary for both variants are compared with the case of undisturbed flow depicted with markers.

\begin{figure}[h!]
\centering
    \includegraphics [scale=0.43]{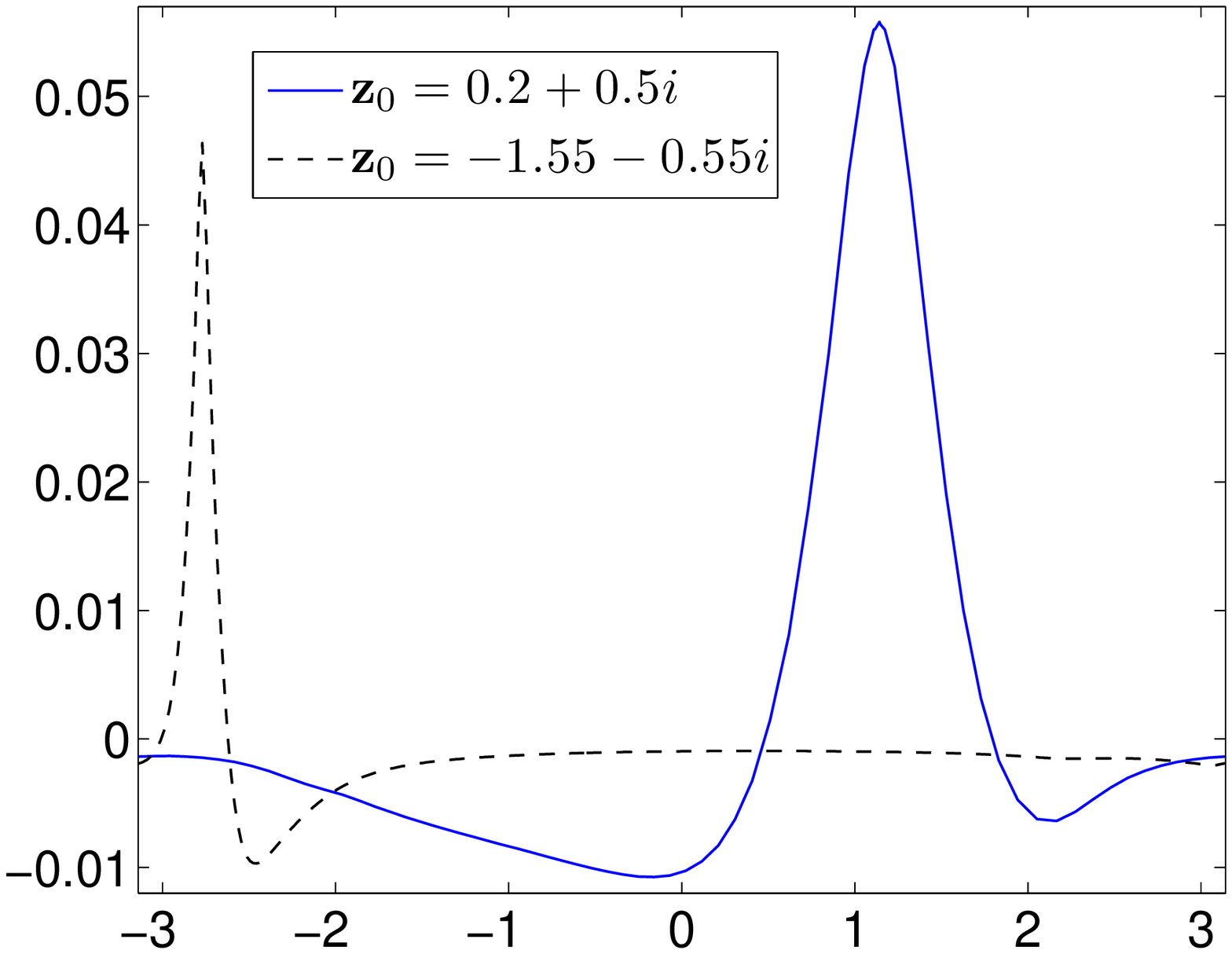}
    \put(-120,0){$\theta$}
    \put(-245,100){$\delta \rho$}
    \caption{Relative deviation of the radius-vector for both variants of inclusion's location from Fig.~\ref{imm_incl_shapes}.
    The reference value of $\rho$ corresponds to the undisturbed flow.}
    \begin{picture}(0,0)(70,100)
    \label{ro_rel_dev1}
    \end{picture}
\end{figure}

Relative deviations of the radius vector, $\delta \rho$, from the one obtained for the undisturbed flow are shown in Fig.~\ref{ro_rel_dev1}. As can be expected,
the maximal distortion of the boundary curve takes place approximately along the direction source-inclusion. Moreover, the shorter the distance between the source and inclusion is, the more pronounced deviation from the reference value is obtained. Since in both cases the source supplies the same volume of fluid in the considered time, one can check the accuracy of computations in terms of the fluid balance. The respective areas are:
\[
A=\frac{1}{2}\int_{-\pi}^{\pi} \rho^2(t_{max},\theta) d\theta.
\]
The relative deviations of $A$ from the benchmark value were: $1.64\cdot 10^{-6}$  and  $4.51\cdot 10^{-5}$ for the first and second location of the obstacle, respectively. We believe that the second value is greater due to the integration error itself, as the relative deviation of $\rho$ has a much sharper maximum in this case (compare Fig.~\ref{ro_rel_dev1}).
However, both obtained results suggest very good accuracy of the solutions as well as very good quality of the Green's function approximation even for relatively large magnitude of the small parameter $\varepsilon$. Note, that accuracy of the uniform asymptotic formula increase (decrease) with time in case of source (sink).

Next, we consider the inclusion with two degrees of freedom (translations) analysing its movement in two cases. In the first of them, the friction term in equations \eqref{Final_system3} - \eqref{Final_system4} is neglected. The second variant accounts for the friction phenomenon. We assume in the computations
that $\kappa=Q/(4\pi/\epsilon)$ and thus the multipliers in both terms representing forces are the same and equal to  $\epsilon Q/(4m)$. In both cases we assume that initial position of the inclusion is ${\bf z}_0(0)=0.1+0.1i$ and its initial velocity is zero.

\begin{figure}[h!]
\centering
    \includegraphics [scale=0.43]{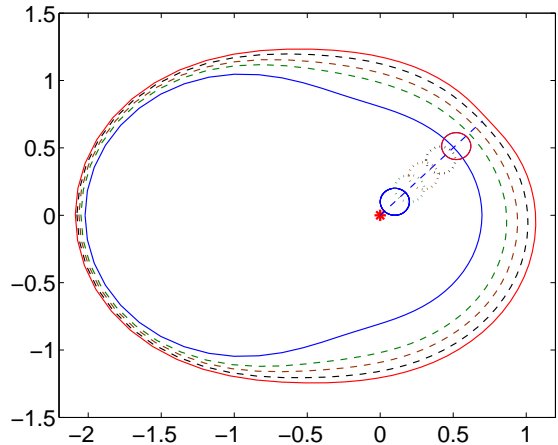}
    %\put(-120,0){$\theta$}
    %\put(-245,100){$\delta \rho$}
    \caption{The domain evolution and obstacle movement for the frictionless variant of the problem. Dashed lines illustrate selected intermediate positions.}
    \begin{picture}(0,0)(70,100)
    \label{incl_mov_shapes}
    \end{picture}
\end{figure}

\begin{figure}[h!]
\centering
    \includegraphics [scale=0.43]{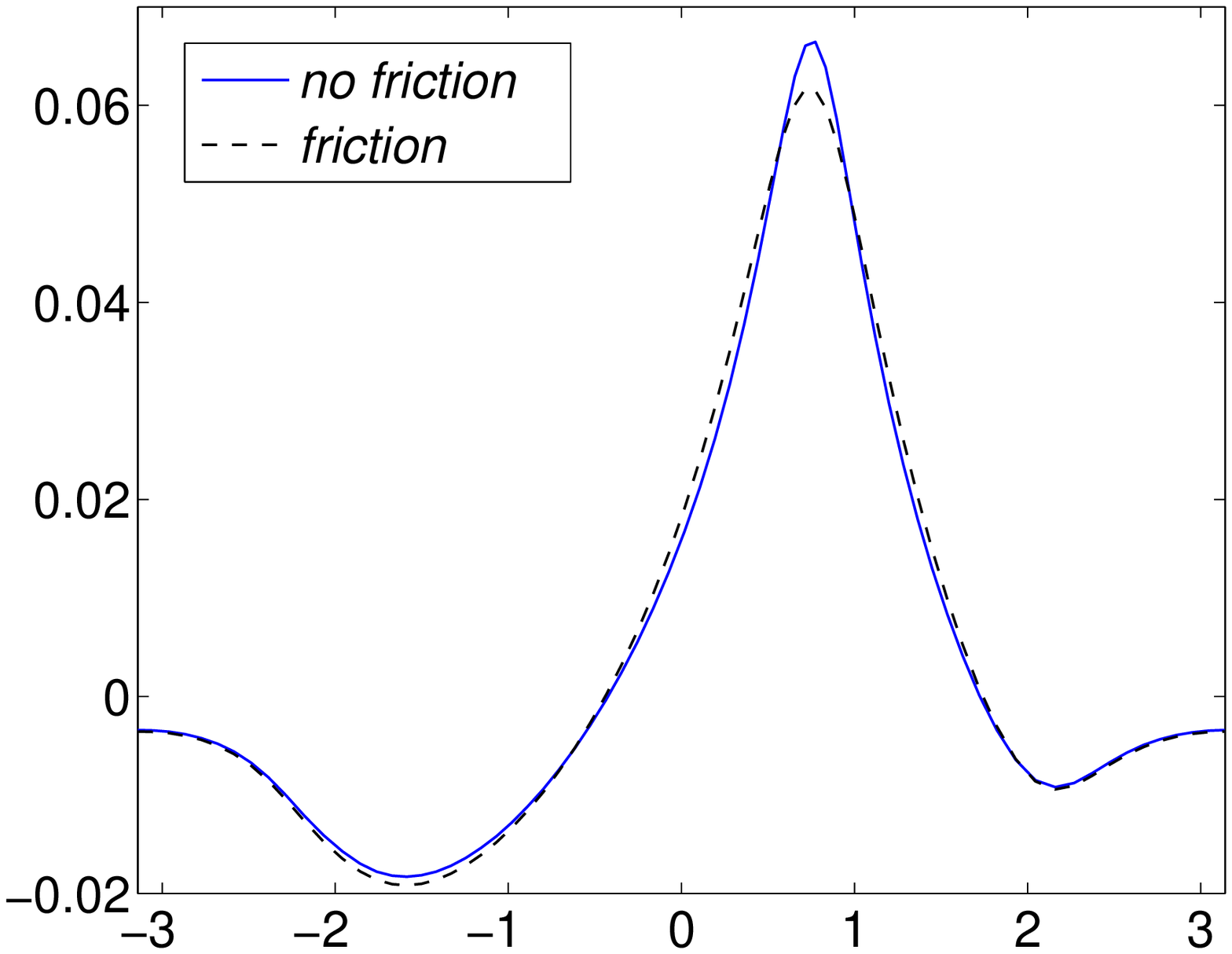}
    \put(-120,0){$\theta$}
    \put(-245,100){$\delta \rho$}
    \caption{Relative deviations of the radius vector from the benchmark value.}
    \begin{picture}(0,0)(70,100)
    \label{ro_rel_dev_mov}
    \end{picture}
\end{figure}

The evolution of the free boundary and the obstacle movement for the frictionless variant are shown in Fig.~\ref{incl_mov_shapes}. Starting from zero initial velocity,
the obstacle moves rectilinearly along the line: the source - center of the inclusion. We do not present a respective picture for the second variant of the problem, as the free boundary shape is hardly distinguishable from the former. Relative deviations from the benchmark values of $\rho (t_{max},\theta)$ for both cases are depicted in Fig.~\ref{ro_rel_dev_mov}.
The balance equation was satisfied this time to the level of $10^{-8}$.

\begin{figure}[h!]
\centering
    \includegraphics [scale=0.43]{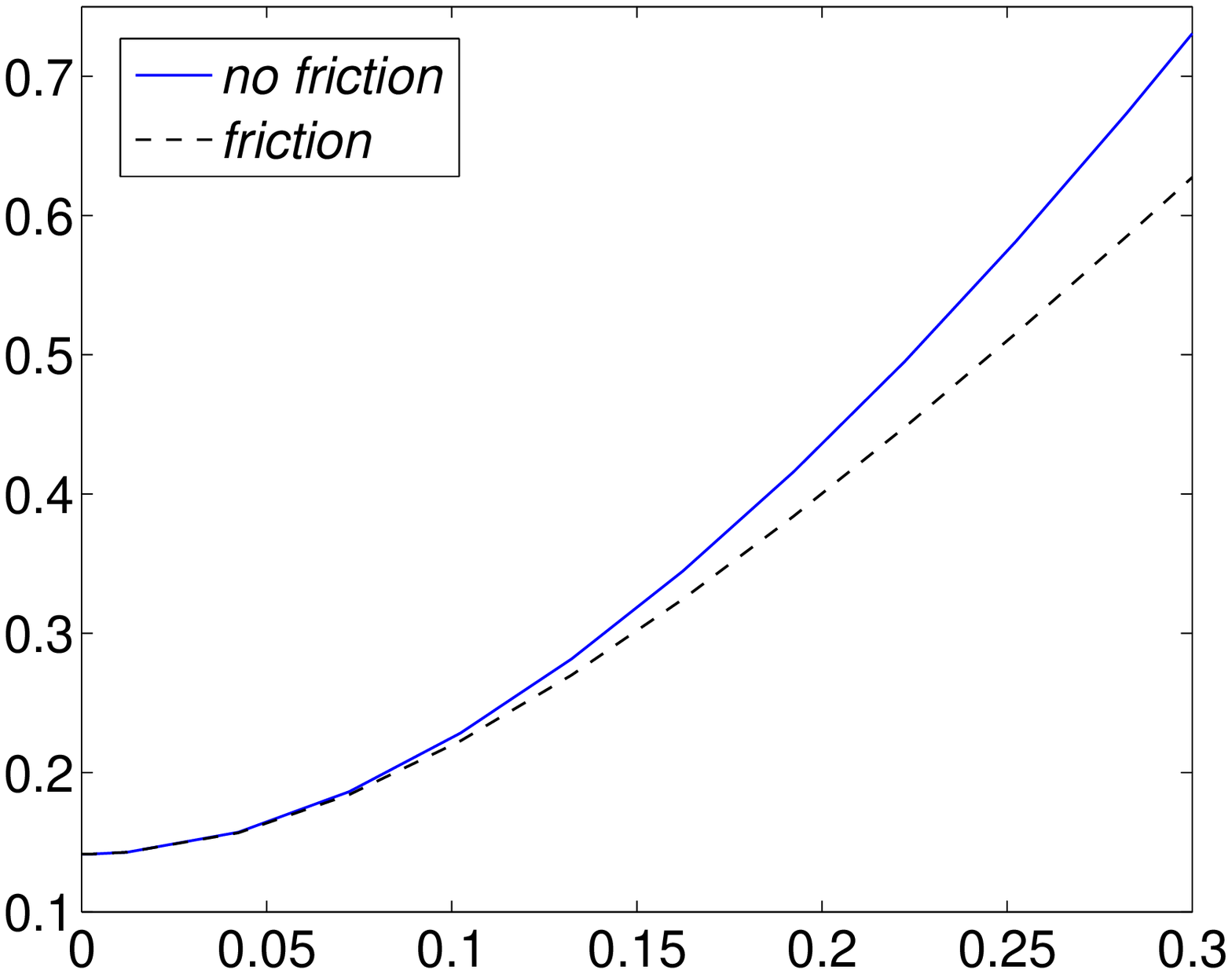}
    \put(-120,0){$t$}
    \put(-245,100){$s(t)$}
    \caption{The distance covered by the inclusion.}
    \begin{picture}(0,0)(70,100)
    \label{s_t}
    \end{picture}
\end{figure}

\begin{figure}[h!]
%M/N=1/300

    \hspace{-2mm}\includegraphics [scale=0.39]{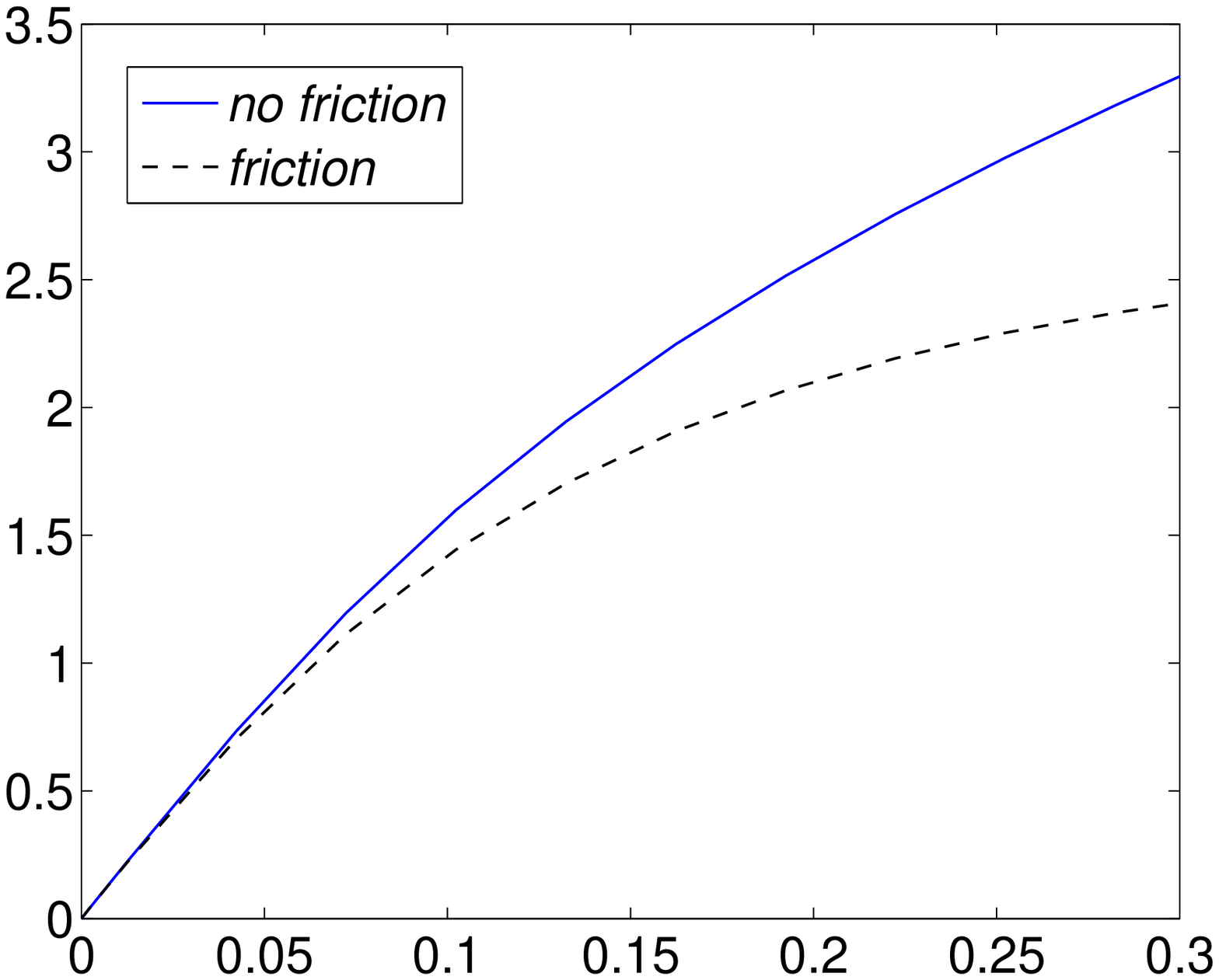}
    \put(-110,-2){$t$}
    \put(-230,90){$|{\bf v}(t)|$}
    \put(-110,-2){$t$}
    \put(-220,150){$\bf{a})$}
%M/N=1/30
    \hspace{-4mm}\includegraphics [scale=0.39]{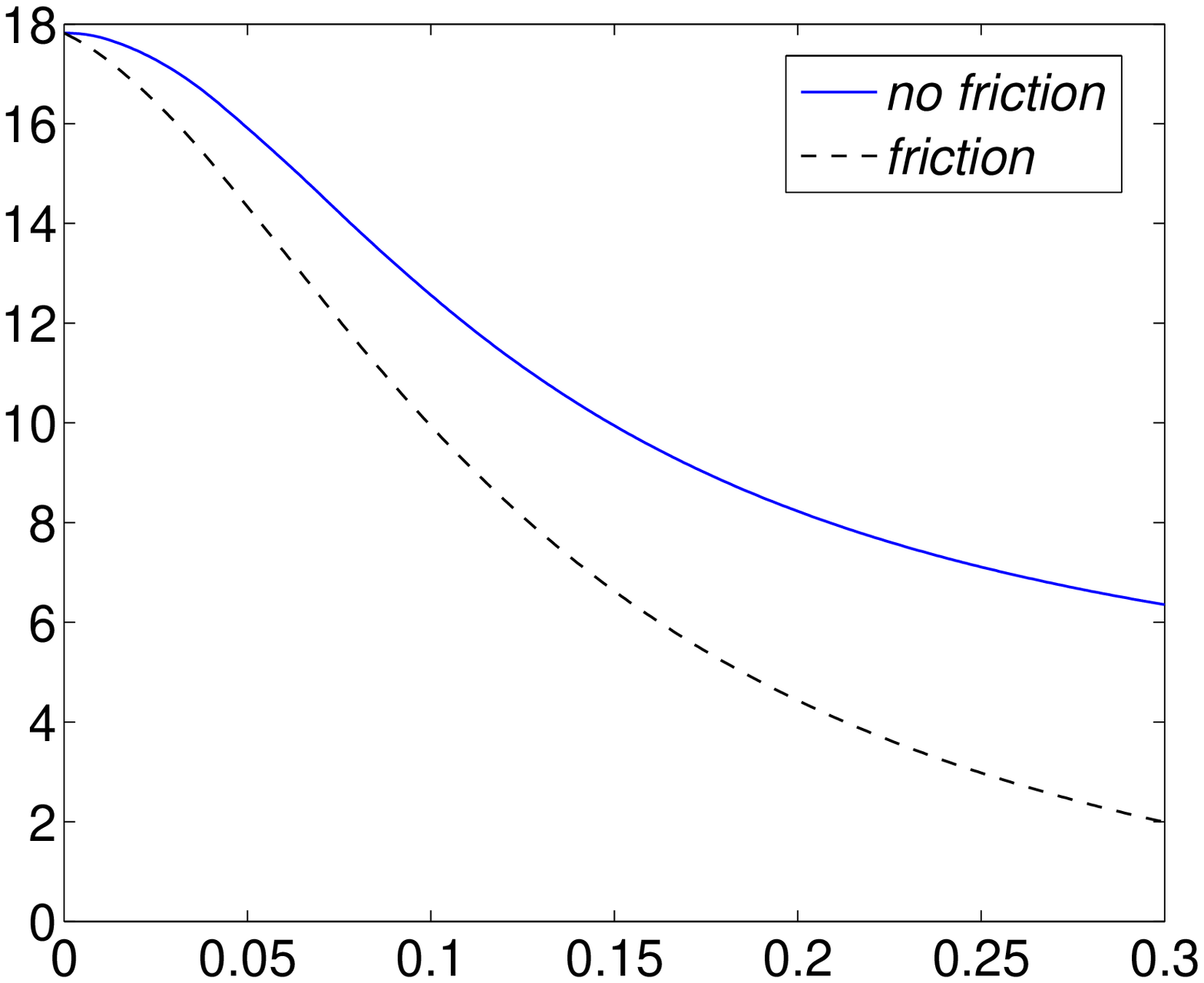}
    \put(-110,-2){$t$}
    \put(-227,90){$|{\bf a}(t)|$}
    \put(-220,150){$\bf{b})$}
    \caption{The absolute values of: a) the inclusion velocity,
    b) the inclusion acceleration.}
    \begin{picture}(0,0)(70,100)
    \label{v_t_a_t}
    \end{picture}
\label{condition}
\end{figure}

The influence of friction on the inclusion movement is shown in Fig.~\ref{s_t}-Fig.~\ref{v_t_a_t}. As the obstacle moves in both cases along the straight line, it is sufficient to present the evolution in time of: the  covered distance (Fig.~\ref{s_t}) -$s(t)$, the absolute value of the velocity (Fig.~\ref{v_t_a_t}a))-$|{\bf v}(t)|$, and the absolute value of the acceleration (Fig.~\ref{v_t_a_t}b)) - $|{\bf a}(t)|$. %{As one can expect, additional friction restricts the particle movement and influences greatly on the free boundary than the frictionless case. }

As anticipated, the influence of friction becomes more pronounced along with the velocity increase, however in the considered time interval it is still far away from making the inclusion movement uniform. Obviously by increasing the value of friction coefficient one can obtain the steady state much faster.

\begin{figure}[h!]
\centering
    \includegraphics [scale=0.45]{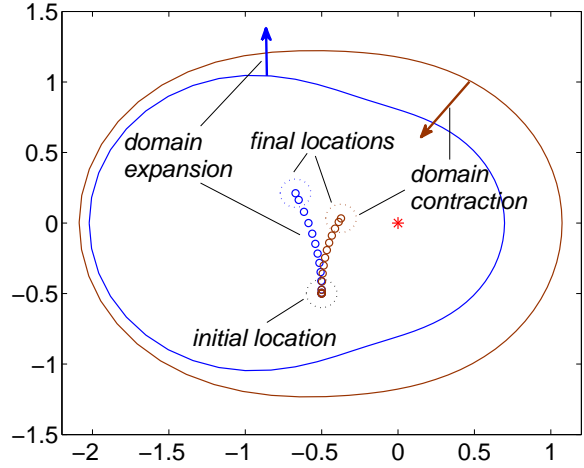}
    %\put(-120,0){$t$}
    %\put(-250,100){$|{\bf a}(t)|$}
    \caption{Domain configuration and obstacle movement. Markers correspond to intermediate positions of the inclusion. Depicted boundary curves define the initial shapes of the domain for respective variants of the problem.}
    \begin{picture}(0,0)(70,100)
    \label{mov_incl_track}
    \end{picture}
\end{figure}

\begin{figure}[h!]
\centering
    \includegraphics [scale=0.45]{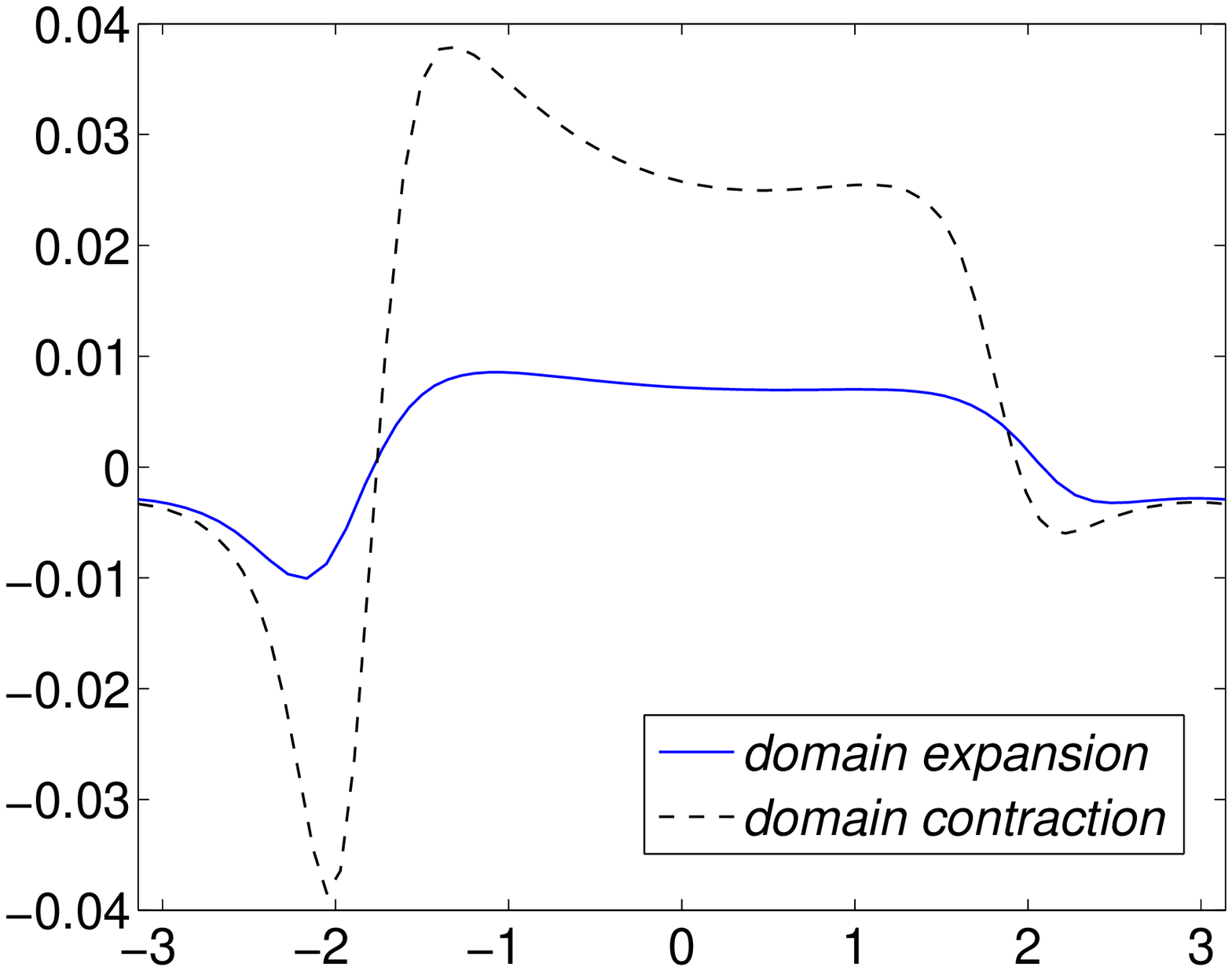}
    \put(-120,0){$t$}
    \put(-260,100){$\delta \rho$}
    \caption{Relative deviations of the radius vector from the benchmark values (without inclusion).}
    \begin{picture}(0,0)(70,100)
    \label{ro_dev_track}
    \end{picture}
\end{figure}

In the last part of our analysis we consider the case when the initial velocity of the inclusion has a non-zero value, and its vector is not collinear
with the line: center of inclusion - source. We investigate the evolution of obstacle track, velocity and acceleration caused by the fluid flow. It is assumed
that the initial position of the obstacle is ${\bf z}_0(0)=-0.5-0.5i$, while its initial velocity yields ${\bf v}(0)=2i$. We consider two variants of the problem depicted in Fig.~\ref{mov_incl_track}. In the first one, the fluid flow is driven by the source and initial shape of the free boundary is described by the internal curve. The second variant assumes the domain contraction caused by the fluid sink. Here, the initial domain is defined by the external curve. This time we shall rather concentrate on the inclusion movement, than on the evolution of the free boundary. The relative deviations of the radius vector from respective reference (benchmark) values are shown in Fig.~\ref{ro_dev_track}. Naturally, the variant with the sink gives more pronounced deformation of the final shape, as the distance between the inclusion and the boundary is much smaller than in the opposite case. The fluid balance equation was satisfied to the level of $10^{-8}$ for the variant of domain expansion, and $10^{-7}$ for domain contraction.

\begin{figure}[h!]
%M/N=1/300

    \hspace{-2mm}\includegraphics [scale=0.39]{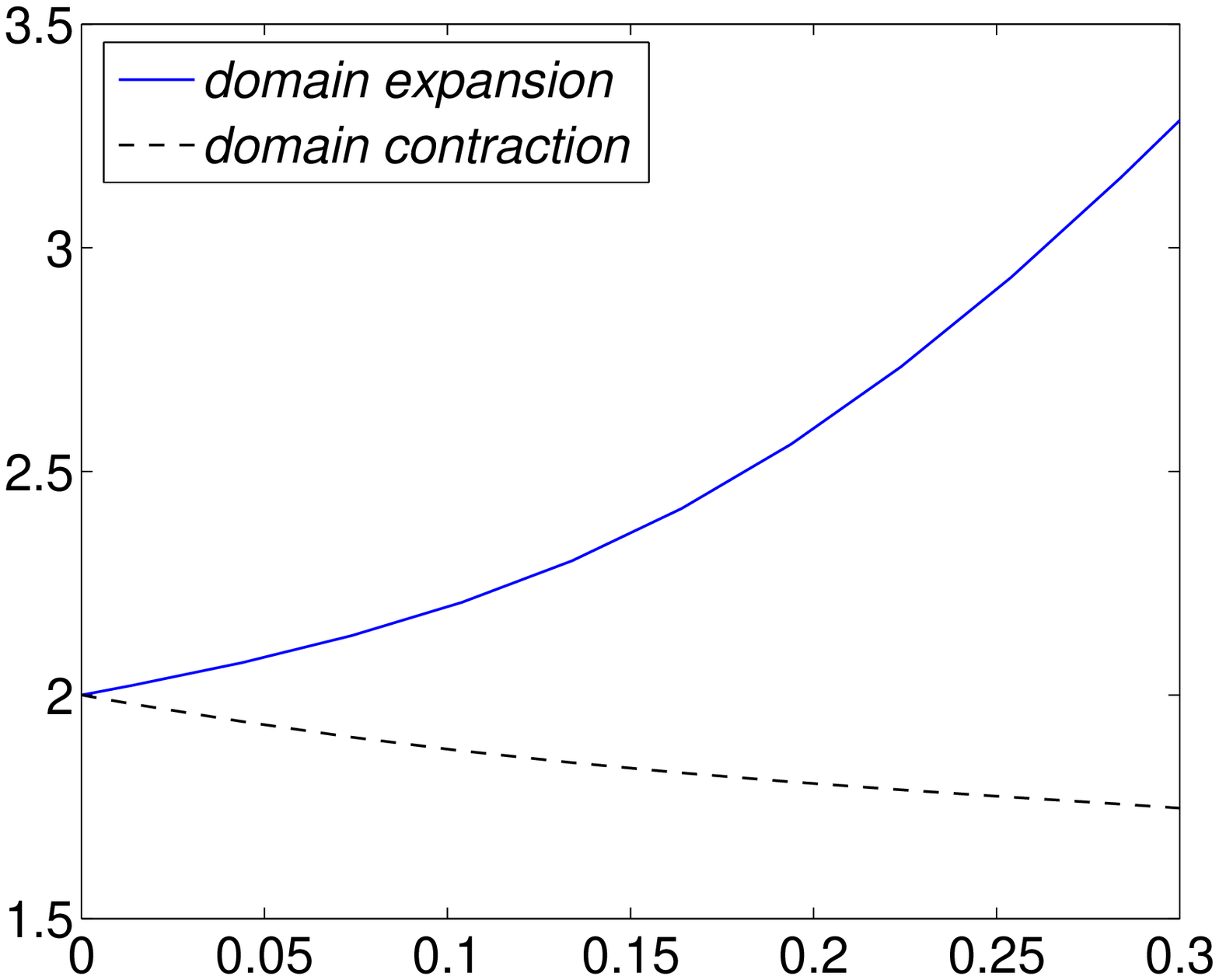}
    \put(-110,-2){$t$}
    \put(-230,90){$|{\bf v}(t)|$}
        \put(-220,150){$\bf{a})$}
%M/N=1/30
    \hspace{-4mm}\includegraphics [scale=0.39]{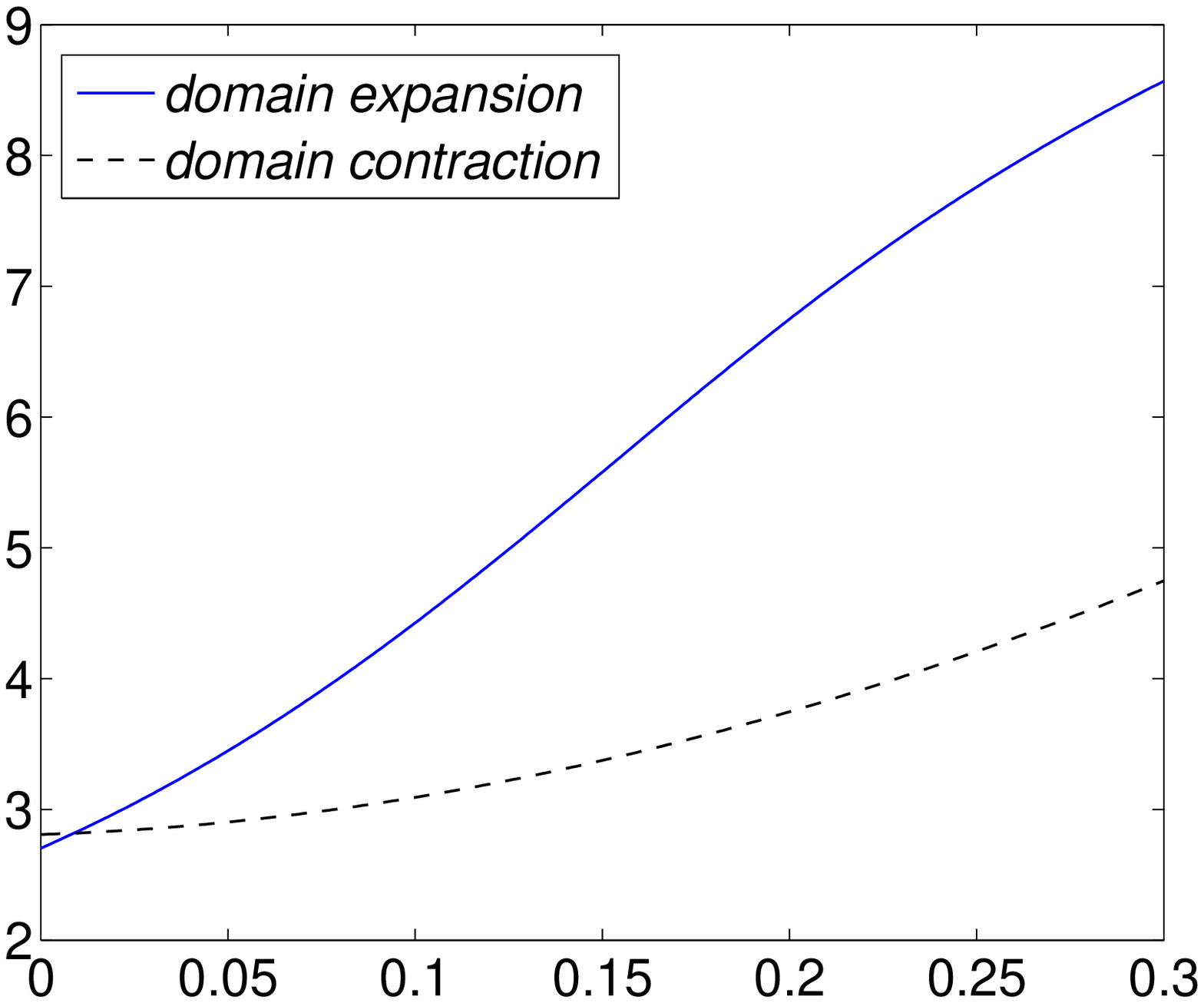}
    \put(-110,-2){$t$}
    \put(-227,90){$|{\bf a}(t)|$}
    \put(-220,150){$\bf{b})$}
    \caption{The absolute values of: a) inclusion velocity,
    b) inclusion acceleration.}
    \begin{picture}(0,0)(70,100)
    \label{a_t_abs}
    \end{picture}
\label{condition}
\end{figure}

The traces of inclusion for both considered cases are shown by markers in Fig.~\ref{mov_incl_track}. It should be emphasized that the imposed initial conditions do not imply kinematic equivalence between both variants of the problem. It is a consequence of different initial accelerations resulting from equations \eqref{Final_system3} - \eqref{Final_system4}. As can be seen in Fig.~\ref{a_t}, Fig.~\ref{a_t_abs}, although the magnitudes of initial accelerations are very close to each other, their vectors directions are almost opposite.

The curvatures of the tracks (bend directions) and the signs of respective components of acceleration are determined by the source/sink activity. In the case of domain expansion the fluid flow direction magnifies the velocity of obstacle. Thus the distance covered is greater than that for fluid sink. Obviously, for other configurations of the initial velocity vector one can expect different trends.

\begin{figure}[h!]
%M/N=1/300

    \hspace{-2mm}\includegraphics [scale=0.39]{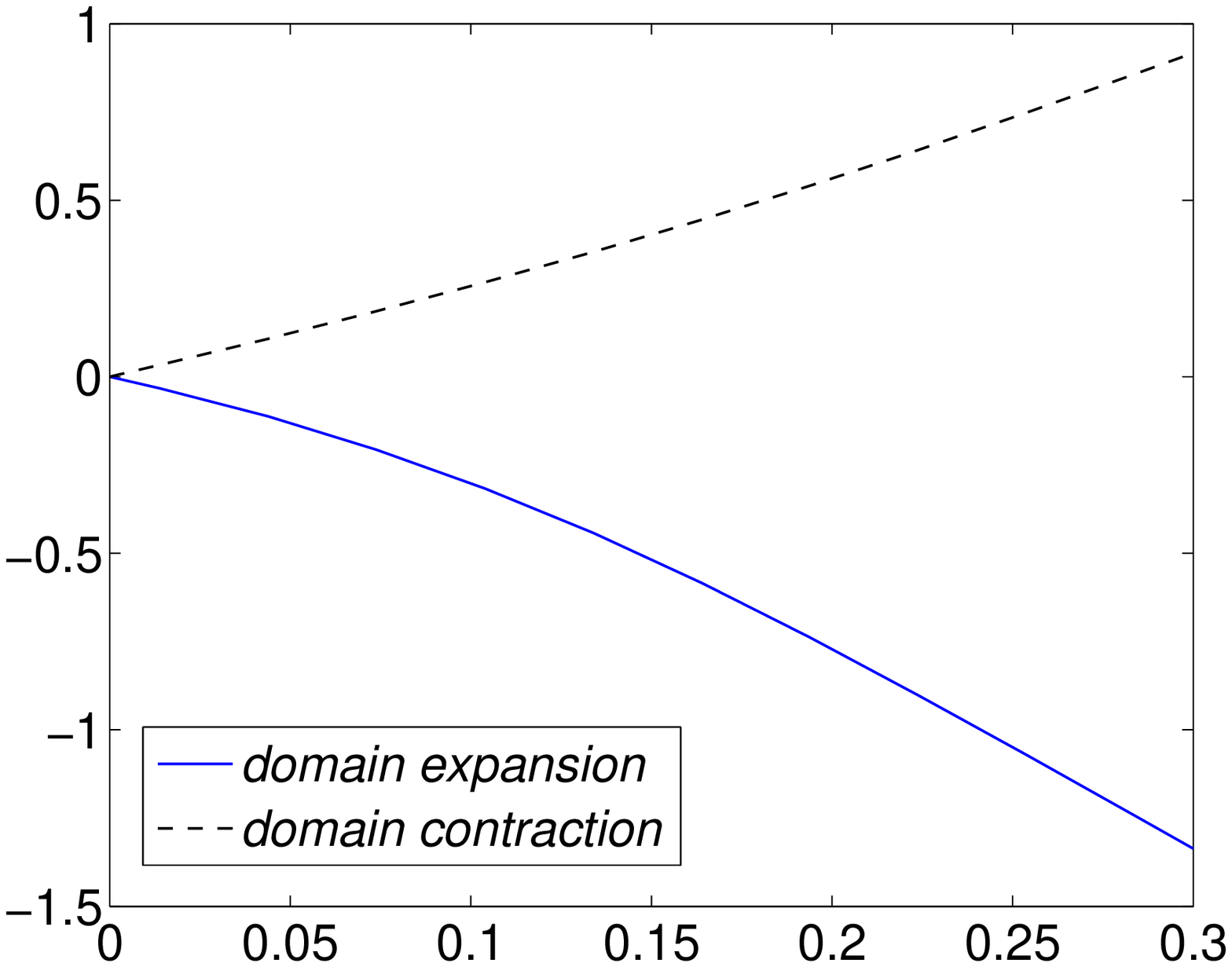}
    \put(-110,-2){$t$}
    \put(-230,90){$v_1(t)$}
    \put(-110,-2){$t$}
    \put(-220,150){$\bf{a})$}
%M/N=1/30
    \hspace{-4mm}\includegraphics [scale=0.39]{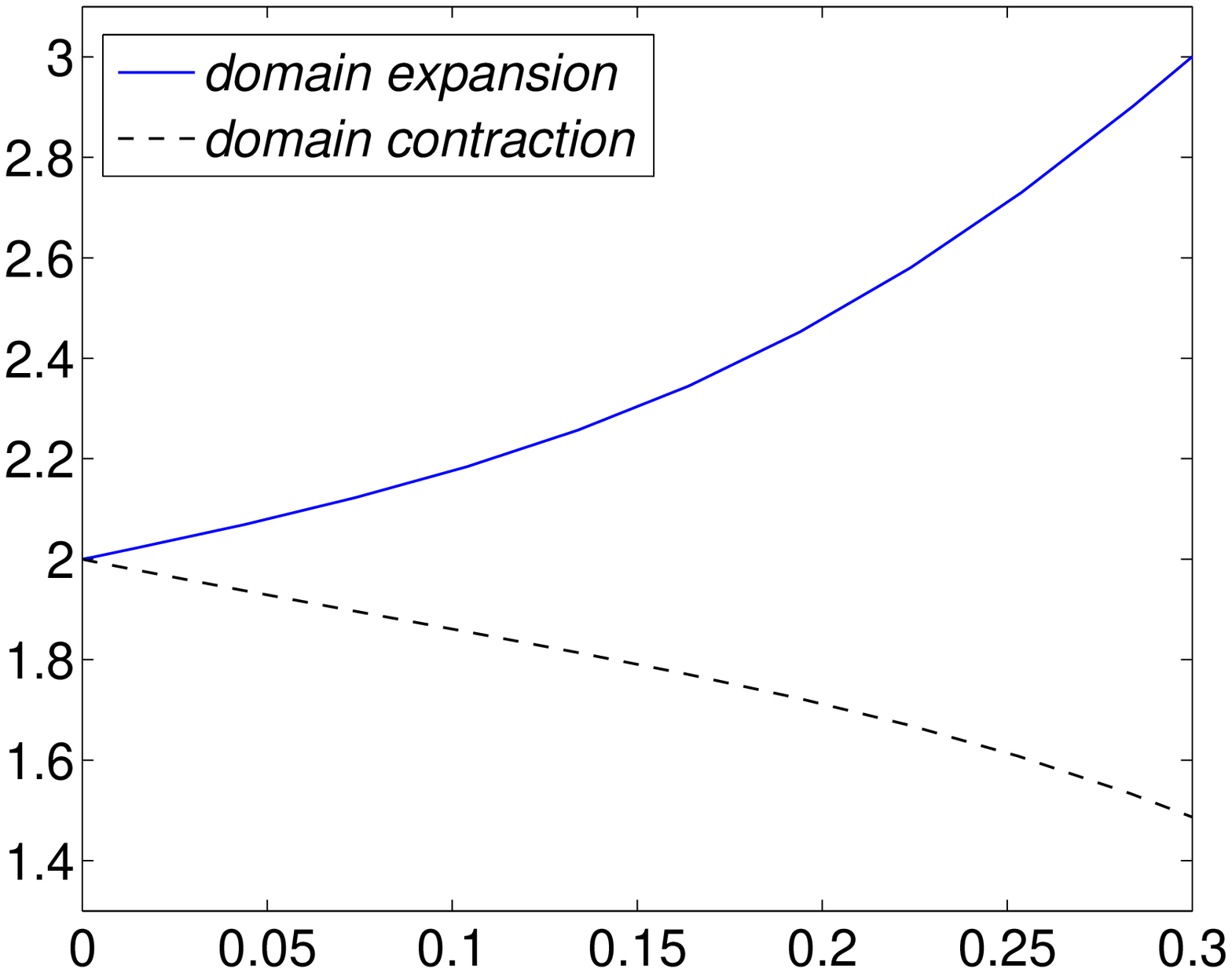}
    \put(-110,-2){$t$}
    \put(-227,90){$v_2(t)$}
    \put(-220,150){$\bf{b})$}
    \caption{Components of the inclusion velocity: a) horizontal,
    b) vertical.}
    \begin{picture}(0,0)(70,100)
    \label{a_t}
    \end{picture}
\label{condition}
\end{figure}

\begin{figure}[h!]
%M/N=1/300

    \hspace{-2mm}\includegraphics [scale=0.39]{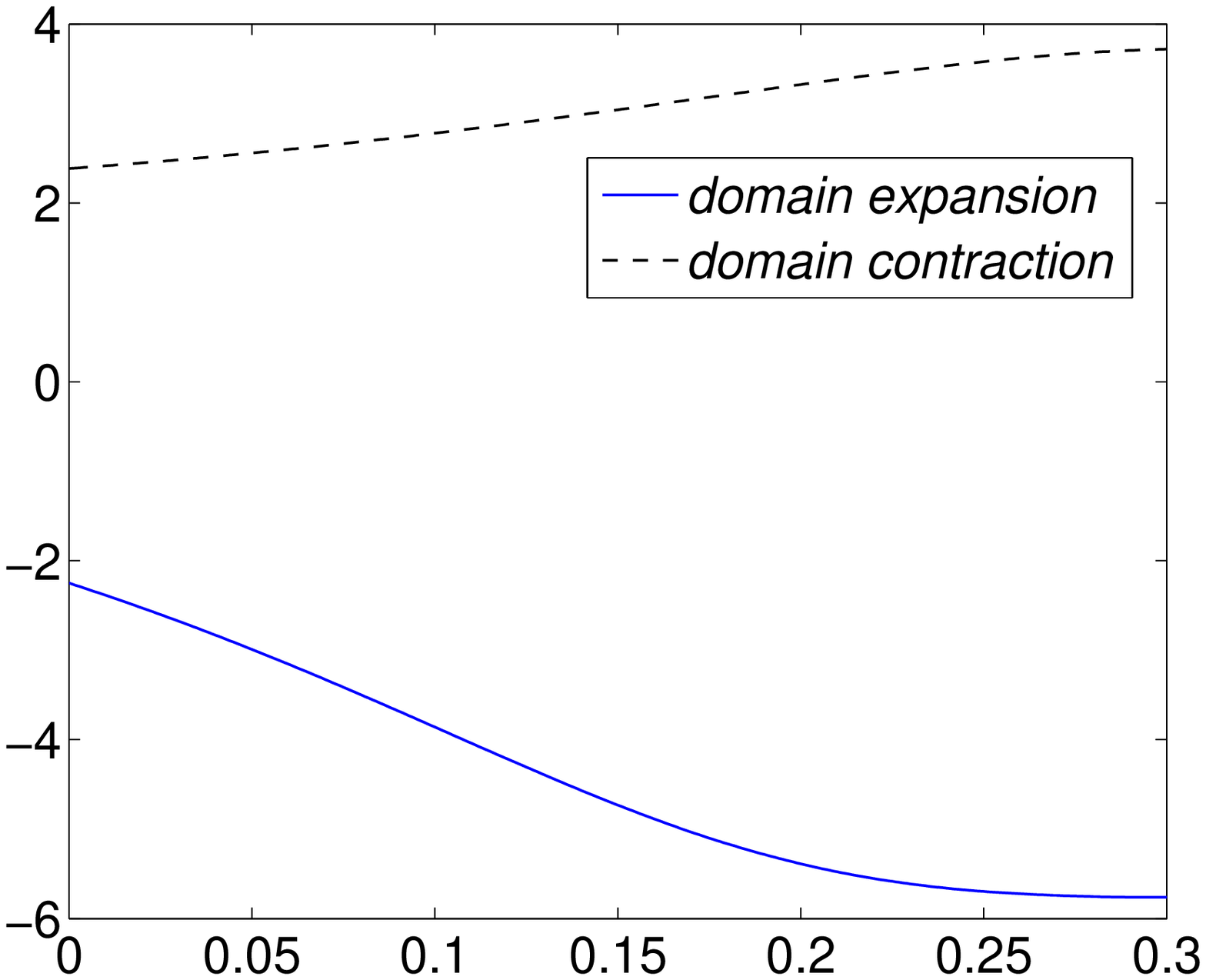}
    \put(-110,-2){$t$}
    \put(-230,90){$a_1(t)$}
    \put(-110,-2){$t$}
    \put(-220,150){$\bf{a})$}
%M/N=1/30
    \hspace{-4mm}\includegraphics [scale=0.39]{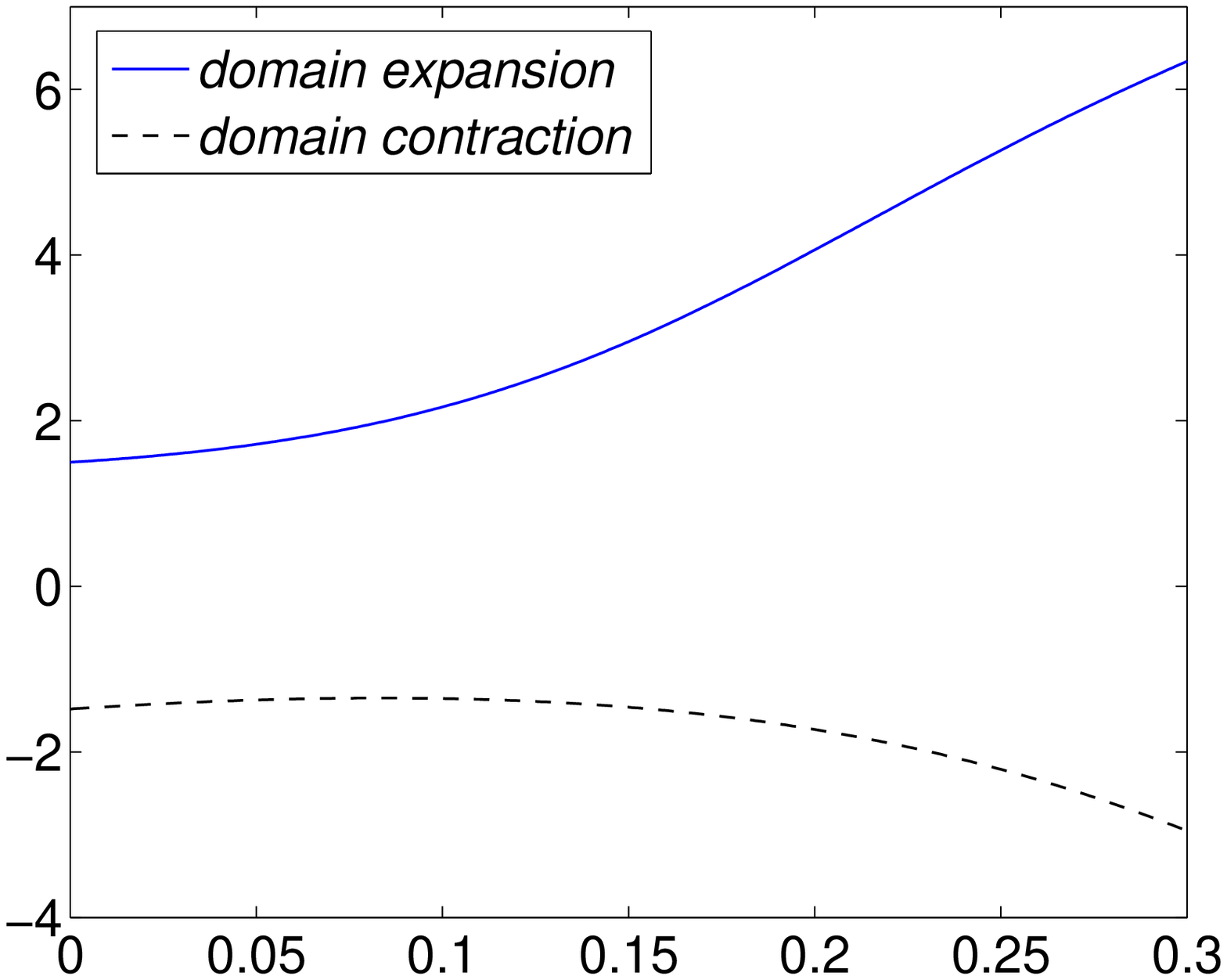}
    \put(-110,-2){$t$}
    \put(-227,90){$a_2(t)$}
    \put(-220,150){$\bf{b})$}
    \caption{Components of the inclusion acceleration: a) horizontal,
    b) vertical.}
    \begin{picture}(0,0)(70,100)
    \label{a_t}
    \end{picture}
\label{condition}
\end{figure}

\begin{figure}[h!]
\centering
    \includegraphics [scale=0.45]{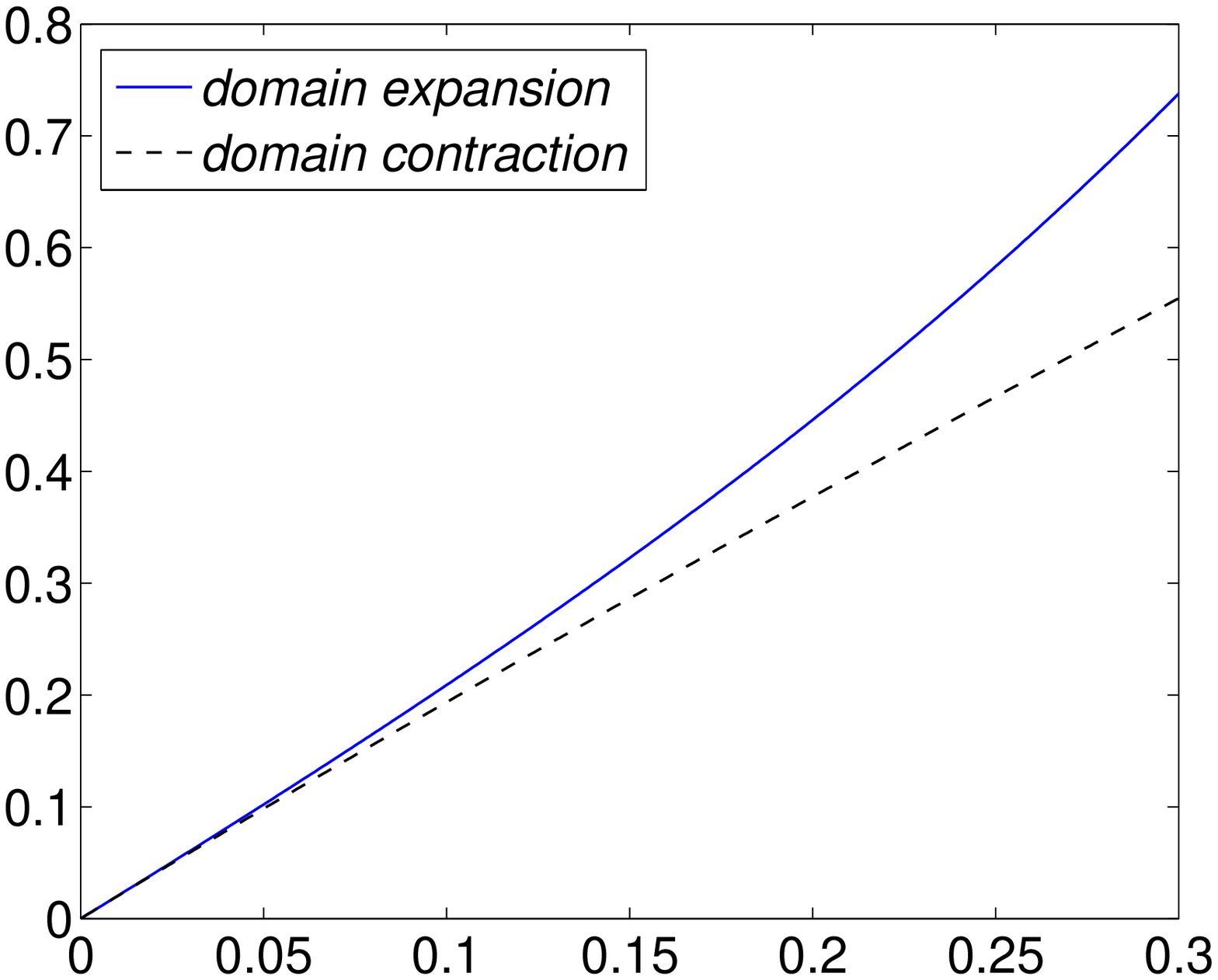}
    \put(-120,0){$t$}
    \put(-260,100){$s(t)$}
    \caption{The distance covered by the inclusion.}
    \begin{picture}(0,0)(70,100)
    \label{s_t_track}
    \end{picture}
\end{figure}

Concluding this section, we have shown that the method utilized the uniform asymptotic expansion for the Green function delivered in \cite{MazMovNie13}
is effective and the computations based on that approach are stable and robust.

%\newpage
\vspace{50mm}
 {\bf Acknowledgement.} The work has been supported by PEOPLE IAPP Project PIAP-GA-2009-251475 HYDROFRAC.

\end{document}